\documentclass[%
reprint,
  aps,prab,
  amsmath,amssymb,
  superscriptaddress
]{revtex4-2}
\usepackage[bottom]{footmisc}
\usepackage{booktabs}
\usepackage{graphicx}
\usepackage{dcolumn}
\usepackage{bm}
\usepackage{hyperref}
\usepackage[mathlines]{lineno}

\newcommand{\KEK}{High Energy Accelerator Research Organization, 1-1 Oho, Tsukuba, Ibaraki 305-0801, Japan}
\newcommand{\IbarakiUniv}{Graduate School of Science and Engineering, Ibaraki University, Mito, Ibaraki 310-8512, Japan}
\newcommand{\UnivofTokyo}{Graduate School of Science, University of Tokyo, 7-3-1 Hongo, Bunkyo-ku, Tokyo 113-0033, Japan}
\newcommand{\SOKENDAI}{The Graduate University for Advanced Studies, Kanagawa 240-0193, Japan}
\newcommand{\CSNS}{China Spallation Neutron Source Science Center, Dongguan 523803, China}
\newcommand{\IHEP}{Institute of High Energy Physics, Chinese Academy of Sciences, Beijing 100049, China}

\begin{document}

\preprint{APS/123-QED}

\title{\textbf{Three-Dimensional Spiral Beam Injection:\\
Design Principles and Experimental Verification
\thanks{Work supported by  JPS KAKENHI Grant Numbers JP19H00673 and JP20H05625.}} 
}%

\author{H.~Iinuma}\email{Contact author: hiromi.iinuma.spin@vc.ibaraki.ac.jp}\affiliation{\IbarakiUniv}
\author{R.~Matsushita}\affiliation{\UnivofTokyo}
\author{M.~A.~Rehman}\thanks{Present address: \CSNS, \IHEP}\affiliation{\SOKENDAI}
\author{H.~Nakayama}\affiliation{\KEK}
\author{S.~Ohsawa}\affiliation{\KEK}
\author{K.~Furukawa}\affiliation{\KEK}

\date{\today}

\begin{abstract}
    A proof of principle experiment of “Three-dimensional spiral beam injection scheme” has been carried out. 
  This injection scheme requires a strongly \textit{x-y} coupled beam to meet magnetic field distribution through solenoid magnet fringe field. 
  In this paper, we introduce outline of experimental setup, results of \textit{x-y} coupling adjustment with DC electron beam of 80 keV. 
	The results of this experiment will be evaluated and improvements for actual operation will be discussed.
\end{abstract}

\maketitle

%


\section{Motivation}

Most elementary particles described by the Standard Model of particle physics (SM) have been discovered through high-energy experiments employing large accelerator facilities with circumferences of several tens of kilometers.
In contrast, trap-based experiments that perform ultra-precision measurements of fundamental properties of nearly stationary elementary particles, as well as protons and neutrons, have actively tested the predictions of the SM with high sensitivity.

However, experimental studies in the energy region where the Lorentz factor $\gamma$ is relatively small are scarce, 
leaving a gap in the systematic examination of the SM from the viewpoint of energy dependence.
This work focuses on the development of an advanced beam-orbit control technique for injecting and storing a charged beam with a few 
$\gamma$ into a compact storage ring, aiming to enable and enhance high-precision measurement experiments in this unexplored energy regime.

Charged elementary particles possess vector properties associated with their quantum-mechanical spin angular momentum, 
characterized by the magnetic (and electric) dipole moments.
This makes them well suited for precision measurements of their interactions with external magnetic (and electric) fields. 
In particular, the spin-precession angular frequency in a magnetic field, $\vec{\omega}$, is given by the Thomas–BMT equation:

\vspace{-2mm}
\begin{equation}
\vec{\omega} 
= \vec{\omega}_{a} + \vec{\omega}_{\eta}
= -\frac{q}{m}
\left[
a_{\mu} \vec{B}
+ \frac{ \eta }{ 2 } \left( \vec{\beta} \times \vec{B} \right)
\right]
\label{eq:t-bmt-e34}
\end{equation}

Here, the particle mass and charge are denoted by $m$ and $q$, respectively, and the velocity normalized by the speed of light 
$c$~is expressed as $\vec{\beta}$.

For example, we focus on the case of the muon here. 
An advantage of using muons is that their mass is approximately 200 times larger than that of the electron, which enhances the sensitivity of the magnetic dipole moment $g$ to quantum corrections from all interactions in the SM—namely, the anomalous magnetic moment ($g$-$2$).
In the above equation, the first term can be written using the muon $g$-$2$ as the anomaly parameter $a_{\mu}\equiv$($g$-2)/2, and we define the corresponding precession frequency as $\vec{\omega}_a$.
The second term corresponds to the component $\vec{\omega}_{\eta}$, which is associated with the parameter~$\eta$ representing the electric dipole moment (EDM).
Within the SM, the EDM is an extremely suppressed quantity, even when quantum corrections are taken into account. Therefore, the observation of a finite EDM value would constitute an immediate discovery of new physics.

The beauty of Equation~\ref{eq:t-bmt-e34} is the second term becomes detectable because it is proportional to the motional electric field experienced by relativistic muons traversing a magnetic field. 
By employing a muon beam with an appropriately tuned energy, Eq.~(\ref{eq:t-bmt-e34}) indicates that two orthogonal angular frequencies,
$\vec{\omega}$, corresponding to the muon $g$-$2$ and EDM, can in principle be separated and measured simultaneously.
Magnitude of the second term with respect to the first term is 1 to 1000 from the upper limit of EDM value reported by the previous experimet (E821~\cite{Bennett:2008dy}).
The precision of the angular-frequency measurement improves with longer observation times of the muon spin precession witn satisfying Eq.~\ref{eq:t-bmt-e34} situation. 
However, muons have a finite lifetime due to the weak interaction (2.2~$\mu$s at rest, decaying into electrons). 
While higher muon energies are advantageous for precision measurements owing to time dilation, 
increasing the beam energy generally requires larger experimental apparatus, which makes difficult to detect vector information of $\vec{\omega}$ from the muon-decay electrons~\cite{Iinuma:2011zz}.

Motivated by these considerations, we conceived a research program to inject and store muons with an energy corresponding to a threefold extension of their lifetime in a compact storage ring with a circumference of 2 m, where systematic uncertainties can be controlled at a level suitable for ultra-precision experiments~\cite{Iinuma:2016zfu},~\cite{Reference_1}.
Based on this concept, we have been advancing the project in a step-by-step manner, supported by multiple competitive research grants.
This paper presents proof-of-principle experiments of the new beam-injection method, the three-dimensional spiral beam injection, proposed in our 2016 NIM paper~\cite{Iinuma:2016zfu}. We report the experimental results and discuss the comparison between the design expectations and the beam parameters achieved in the actual beamline. The introduction is given below.

\begin{figure}[htb]
 \centering
	 \includegraphics[width=\linewidth]{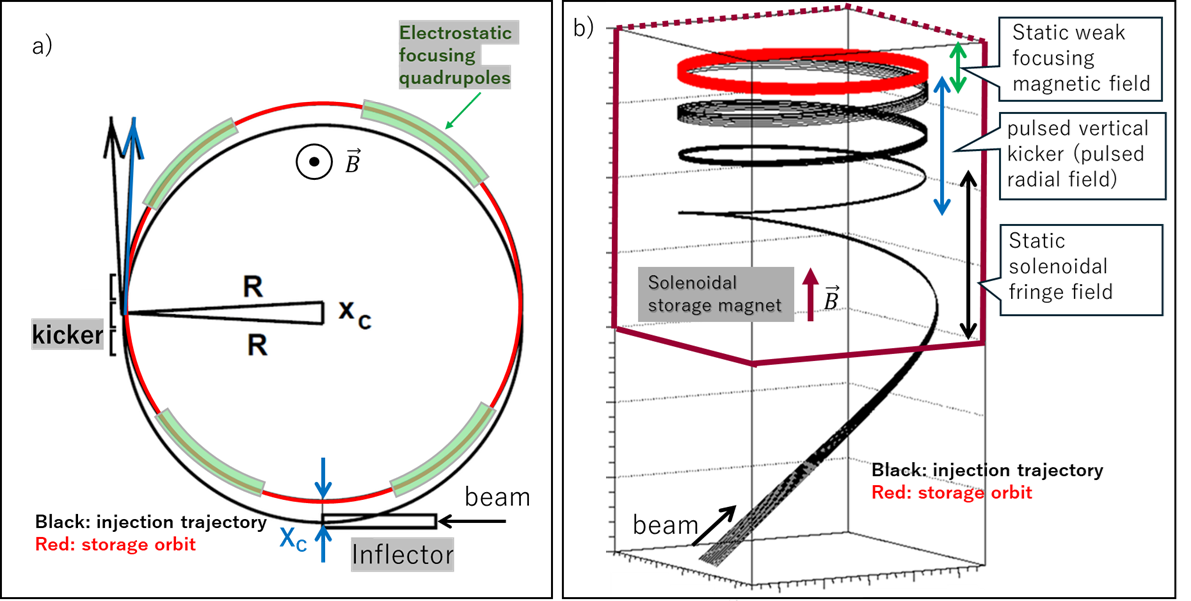}
 \caption{Conventional beam injection vs. 3-D injection scheme. 
	 }
      \label{fig:2-D_3-D}
 \end{figure}

A conventional beam injection is typically realized by magnetically steering the beam 
from a transport line (straight section) into the storage orbit inside a high-field magnet using devices 
such as septum magnets or inflectors~\cite{ipac2018-inflector}. 
A schematic illustration of this conventional injection scheme is shown on the left side of Fig.~\ref{fig:2-D_3-D}.
In contrast, for the injection of a relativistic-energy beam into a compact storage ring with a sub-meter diameter, 
as discussed in this paper, we have developed a novel approach based on a three-dimensional spiral injection trajectory 
inside a solenoidal magnet, rather than applying existing injection techniques.
As shown on the right side of Fig.~\ref{fig:2-D_3-D}, 
this new method actively exploits the fringe magnetic field distributed vertically with respect to the orbital plane for beam-orbit control, 
enabling beam injection without the need for external devices that generate additional electromagnetic fields, such as septa or inflectors.

Furthermore, because this method does not require external devices near the storage magnet—which could otherwise introduce magnetic field imperfections—it allows the magnetic field in the storage region to be locally tuned with sub-ppm precision.
Notably, the injection is realized solely through magnetic-field control generated by currents, making this approach a distinct injection technique based solely on magnetic-field control.

A proof-of-principle experiments to demonstrate its feasibility is discussed in this paper. 
In this experiment, visualization of the three-dimensional spiral trajectory was set as a key milestone. A DC beam from an electron gun was injected into nitrogen gas, and the ionization-induced light emission along the beam path was evaluated in a semi-online manner while optimizing the beam phase-space tuning.
As discussed in Sec.~\ref{sec:SITE}, the experimental setup has several limitations because of the placement.
Rather than simply demonstrating operation consistent with the design, this paper discusses practical strategies for beam tuning and methods for evaluating beam phase-space quality, validated through experimental beam data.
The present paper focuses on the tuning of the coupled $x$–$y$ phase space, while the experimental demonstration of beam storage obtained in the same campaign is presented in a companion paper by Matsushita et al.~\cite{matsuPRL}.

\section{Kinematics of the Three-Dimensional Spiral Beam Injection}
\label{sec:gaiyou}
In this section, we discuss the kinematics of the three-dimensional spiral beam injection, 
the method for deriving the required phase-space parameters at the injection point of the solenoidal storage magnet, and the technique for suppressing beam divergence along the solenoid axis inside the storage magnet.~\cite{OideBeamDynamics},~\cite{SAD}
The discussion is developed while presenting concrete numerical values of beam parameters related to the demonstration experiment of three-dimensional spiral beam injection. The underlying principle, however, is the same as that of the three-dimensional spiral injection designs currently being developed for projects such as the muon beamline at J-PARC. Studies on those designs are summarized elsewhere~\cite{IPAC2023,Iinuma:ipac2025-wepm029,Iinuma:ipac2024-wepg46,Reference_3,Ogawa:ipac2024-thad1,Ogawa:ipac2025-wepm055}.

\begin{figure}[htb]
 \centering
	 \includegraphics[width=\linewidth]{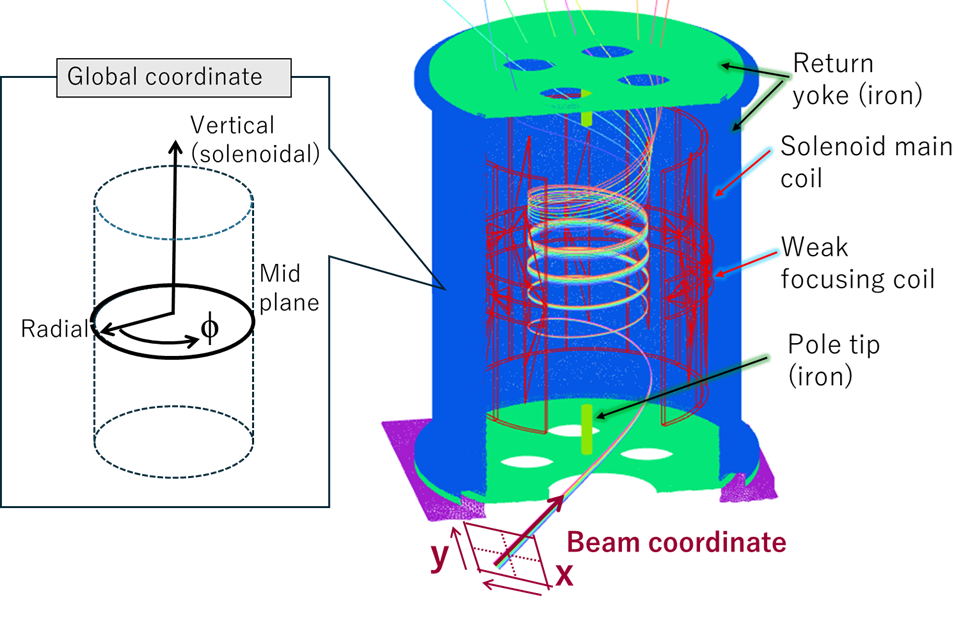}
	 \caption{Image of the 3-D spiral beam injection for this demonstration experiment.
	 OPERA-3D model~\cite{opera} and flat-shaped beam samples are shown.
	 }
      \label{fig:miniSolOPERA}
 \end{figure}

Throughout this paper employs two coordinate systems, depending on the context.
One is a cylindrical coordinate system with the origin at the center of the solenoid magnet, and the other is a beam-based coordinate system with the beam trajectory taken as the reference origin.
The solenoid axis is defined as the vertical direction, the transverse direction as radial, 
the azimuthal angle along the circulating direction as $\phi$, and the injection pitch angle with respect to the plane normal to 
the solenoid axis as $\theta$.
When necessary, we also use a beam-coordinate system defined along the design trajectory for beam-optics considerations.
A schematic illustration of the three-dimensional spiral beam injection is shown in Fig.~\ref{fig:miniSolOPERA} together with an explanation of the coordinate systems.

 \subsection{Derivation of Axis-Symmetric Correlations (X–Y Coupling) Using a Flat Beam}

A beam propagating along a three-dimensional spiral trajectory in an axis-symmetric solenoidal magnetic field can be decomposed into cyclotron motion (circulation in the $\phi$ direction) and motion along the solenoid axis.
The former experiences a centripetal force arising from the cross product of the magnetic flux density along the solenoid axis and the beam momentum, resulting in a continuous radial focusing force.

The latter motion requires appropriate shaping of the beam phase space in accordance with the spatial distribution of the fringe magnetic field of the storage magnet, specifically the radial magnetic-field component $B_R$.

Since the radial magnetic-field component $B_{R}$ depends on both the vertical and radial coordinates, the integrated radial magnetic field 
$\langle B_RL \rangle$~experienced by a single particle traveling at a speed $v$
 along the reference trajectory in the magnetic field can be expressed as

\begin{equation}
	<B_{R}L> = \oint B_R(\vec{r}) ds,~ds=v dt  
\label{eq:eq2}
\end{equation}

Here vectors of a single particle's position and velocity in the global coordinate in fig.~\ref{fig:miniSolOPERA} are given by
\begin{eqnarray}
\vec{r}&=&(R \cos\phi, vertical, R\sin\phi), \\ \nonumber
    \vec{v}&=&(v \cos\zeta \cos\psi,v \sin\zeta, v \cos \zeta \sin \psi),\\ \nonumber
    \cos{\theta}&=&\frac{Rv\cos{\zeta}\cos{(\phi-\psi)}+vertical\cdot~v\sin{\zeta}}{|r||v|}.
    \label{eq:eq3}
\end{eqnarray}

The correlation required at the injection point is essentially determined by the integrated $\langle B_RL \rangle$ value from the injection point to the storage region inside the magnet.
Beam injection into an axially symmetric solenoidal magnetic field therefore requires a correlation between the longitudinal and radial components of the motion—what is commonly referred to as \textit{x–y coupling}. In this paper, we adopt this terminology, where the \textit{x–y coupling} specifically refers to the correlation defined in the beam-coordinate system at the injection point, as illustrated in fig.~\ref{fig:2-D_3-D}.
Given this definition, the most straightforward and logically consistent way to determine the required correlation is to examine the backward-tracked trajectories that originate in the storage region and are propagated upstream to the injection point.
Because the backward-derived correlation exhibits a slight dependence on the choice of reference trajectory, a fine adjustment is necessary.
Consequently, a central task of the injection design is to identify a reference trajectory for which the \textit{x-y} correlation is minimally sensitive to this dependence.
    
More concretely, the phase-space distribution of the beam at the injection point should be designed such that the integrated magnetic field $\langle B_RL \rangle$ experienced by individual beam particles along their respective injection trajectories is as uniform as possible, 
despite their finite spatial and angular spreads.

 \begin{figure}[htb]
 \centering
	 \includegraphics[width=0.5\linewidth]{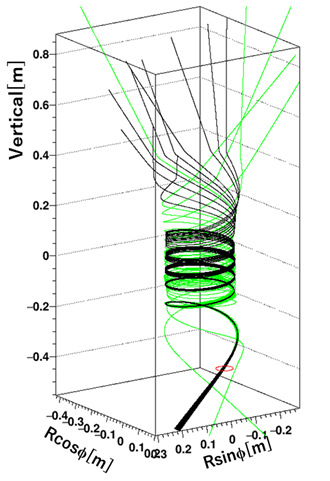}
	 \caption{
	      Black: Three-dimensional spiral trajectories with flat $x$–$y$ correlation, corresponding to the type-1 configuration shown in Fig.~\ref{fig:ribbon_soukan}. 
Green: Trajectories with opposite $x$–$y$ correlation, referred to as type-2. Red circle in the figure denotes entrance of storage magnet as shown in fig.~\ref{fig:miniSolOPERA}.
A clear divergence along the solenoid axis is observed in this case.}
      \label{fig:A1_A1Reverse_3D}
 \end{figure}
 \begin{figure}[htb]
 \centering
	 \includegraphics[width=0.7\linewidth]{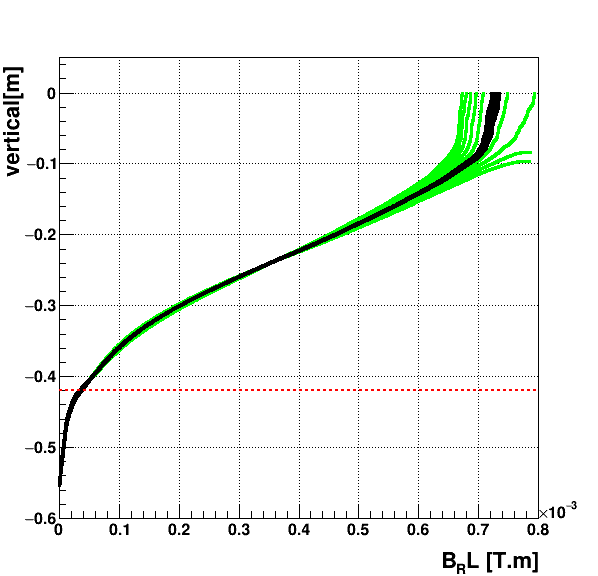}
	 \caption{Vertical diverging as a function of $B_RL$ along each trajectory.  
	 Beam samples are same as Fig.~\ref{fig:A1_A1Reverse_3D}.The red dashed line indicates the bottom surface of the iron yoke of the storage solenoid magnet (see fig.~\ref{fig:miniSolOPERA}).
     As the beam spreads in the vertical direction for $\mathrm{vertical}>$~-0.2m in Fig.~\ref{fig:A1_A1Reverse_3D}, the corresponding 
     $\langle BrL \rangle$~distribution shown here also broadens.}
      \label{fig:BrL}
 \end{figure}

By satisfying this condition, beam divergence along the solenoid axis can be effectively suppressed, enabling stable beam control during injection.

\begin{figure}[htb]
 \centering
	 \includegraphics[width=\linewidth]{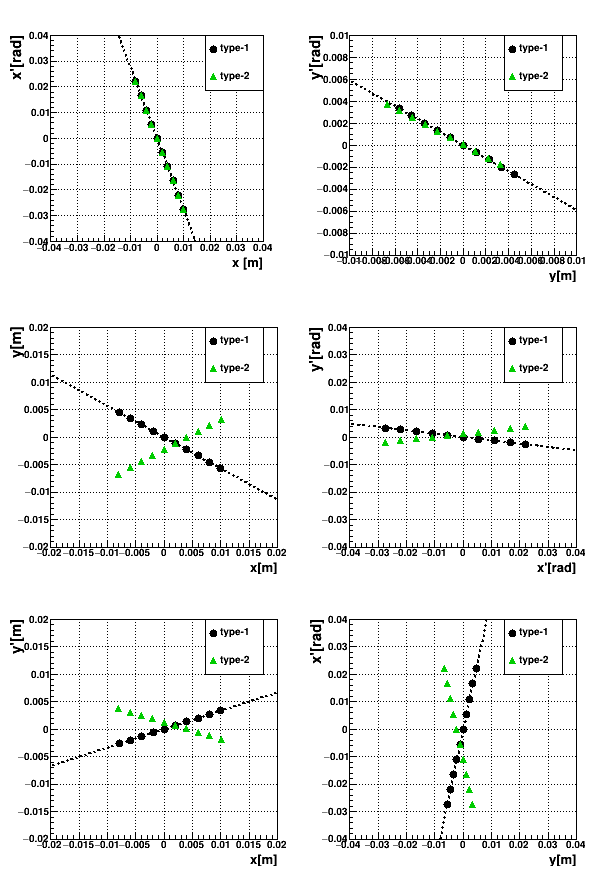}
	 \caption{Strong correlation of flat-shaped beam samples at the injection point(vertical=-0.55~m in fig.~\ref{fig:A1_A1Reverse_3D}).~These strong correlations result from the axisymmetric solenoidal field. Black circles (green triangles) are as in ten black (green) trajectories in fig.~\ref{fig:A1_A1Reverse_3D}
	 }
      \label{fig:ribbon_soukan}
 \end{figure}

The beam phase space at the injection point shown in fig.~\ref{fig:miniSolOPERA} is projected onto the
\textit{x-y} plane of the beam coordinate system and illustrated in fig.~\ref{fig:ribbon_soukan}.
To clearly demonstrate the correlations, we consider an idealized flat (ribbon) beam composed of ten representative trajectories around the reference orbit, neglecting the beam emittance.

Figure~\ref{fig:ribbon_soukan} shows the projection of the spatial and angular spreads of the ribbon beam onto the \textit{x-y} plane in the beam coordinate system as introduce in fig.~\ref{fig:miniSolOPERA}.
The fact that the correlation slopes $x$-$x'$, $y$-$y'$, $x$-$y$, $x$-$y'$, $y$-$x'$ are all negative indicates that the beam is in a focusing condition for both normal and skew components.
The corresponding numerical values are summarized in Table~\ref{tab:soukan}.

\begin{table}[!hbt]
   \centering
	\caption{Slope correlation of type-1 in fig.~\ref{fig:A1_A1Reverse_3D} and fig.~\ref{fig:ribbon_soukan}.}
   \begin{tabular}{lcc}
       \toprule
	   correlation  & slope values &note \\
	    pattern &  m/rad or rad & \\
       \midrule
	   x-x'   &      -2.747 &normal focus   \\ 
	   y-y'   &     -0.550 &normal focus   \\ 
	   x-y    &      0.555 &  \textit{x-y coupling} \\ 
	   x'-y'  &      0.111 &   \textit{x-y coupling}\\
	   x-y'   &      -0.306 &skew focus    \\
	   y-x'   &  -4.946 &skew focus    \\
       \bottomrule
   \end{tabular}
   \label{tab:soukan}
\end{table}

Figure~\ref{fig:A1_A1Reverse_3D} shows the reference orbit together with the results of the three-dimensional spiral injection for ten trajectories (black) that are given appropriate correlations with respect to the reference orbit, as listed in Table~\ref{tab:soukan}.
The trajectories shown in green correspond to the case where the correlation parameter
$x$-$y$ is inverted to a negative value, while keeping the correlations $x$-$x'$ and $y$-$y'$ unchanged. 
As a consequence, the parameters $x$-$y'$ and $y$-$x'$ become positive, leading to a defocusing skew component, and the integrated magnetic field
$\langle B_RL \rangle$ along each trajectory diverges during the injection process.
As a result, the trajectory group shown in green exhibits a larger spread along the solenoid axis compared to the black trajectory group.

Figure~\ref{fig:BrL} presents a comparison of the integrated $\langle B_RL \rangle$ inside the storage magnet between 
the ten trajectories with the ideal correlations (black) and those with the reversed sign of \textit{x-y} (green).
The red dashed line indicates the bottom surface of the iron yoke of the storage solenoid magnet (see Fig.~\ref{fig:miniSolOPERA}). 



\begin{figure}[htb]
 \centering
	 \includegraphics[width=\linewidth]{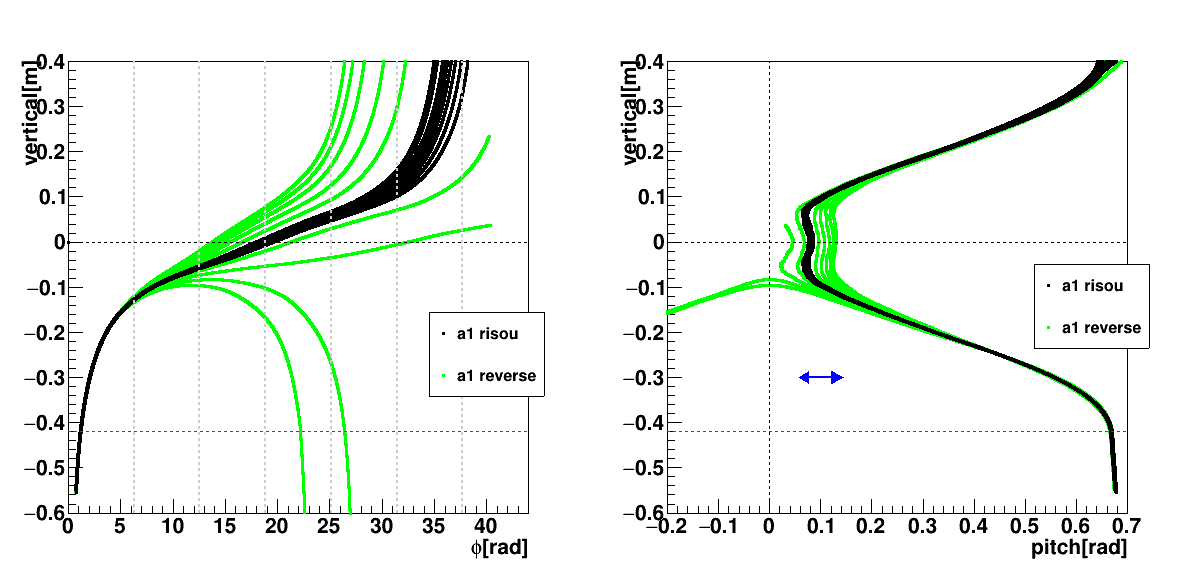}
	 \caption{
     Two plots in which the vertical position is plotted as a function of the azimuthal angle~ $\phi$~(left) and the injection angle~$\zeta$~(right).
	 }
      \label{fig:A1_A1Reverse}
 \end{figure}

The broadening of $\langle B_RL \rangle$ for the green trajectories at vertical $\geq -0.15$ m is consistent with Fig.~\ref{fig:A1_A1Reverse}, where a pronounced vertical divergence appears after one turn following injection, leading to a larger spread in the injection angle $\zeta$ within the vertical range of $\pm 0.1$ m.

Not only the slope of the \textit{x–y} correlation but also deviations in the
$x$–$x'$ and $y$–$y'$ slopes from the ideal backward-tracked values naturally
lead to a divergence of the integrated $\langle B_R L \rangle$.
Although the cross-plane slopes $x$–$y'$ and $y$–$x'$ formally govern this
divergence, practical beamline tuning is mainly achieved through the control
of a combination of $x$–$x'$, $y$–$y'$, and $x$–$y$ correlations, while the
other components are adjusted only indirectly.
Further details are provided in Appendix Sec.~\ref{sec:X-Ysoukan}.

\subsection{Determination of Twiss Parameters Satisfying the Required X–Y Correlations}
\label{sec:soukan-twiss}

To generate a beam distribution with finite emittance that satisfies the required \textit{x-y} correlations based on fig.~\ref{fig:ribbon_soukan} 
and Table~\ref{tab:soukan}, it is necessary to determine the corresponding Twiss parameters.
To make this strategy concrete, we prepared three representative sets of Twiss parameters at the injection point. Type-(A) is constructed directly from the \textit{x-y} correlation obtained by backward tracking from the storage region, whereas type-(B) and type-(C) are obtained by intentionally varying the correlation around this reference solution. These three optics settings are summarized in Table~\ref{tab:twiss}.

\begin{table}[!hbt]
   \centering
   \caption{Twiss parameters and area of guidline}
   \begin{tabular}{lccc}
       \toprule
	    symbol&   type-(A) & (B)  &(C)\\
       \midrule
	      $\alpha_x$~[m/rad] &  27.38 &26.83 &26.61 \\ 
	      $\alpha_y$~[m/rad] &  5.61 &6.38 &5.43\\ 
	      $\beta_x$~[m]   &  9.91 &9.42 & 10.22 \\ 
	      $\beta_y$~[m]   &  9.79 &10.75 & 9.94 \\ 
	      $r_1$ &    7.68 &    7.49&    8.12 \\ 
              $r_2$ &    3.03 &    2.96&    3.28 \\
	      $r_3$ &    3.05 &    3.54&    2.83 \\
	      $r_4$ &    0.97 &    1.23&    0.85 \\
       \bottomrule
   \end{tabular}
   \label{tab:twiss}
\end{table}

\begin{figure}[htb]
 \centering
     \includegraphics[width=\linewidth]{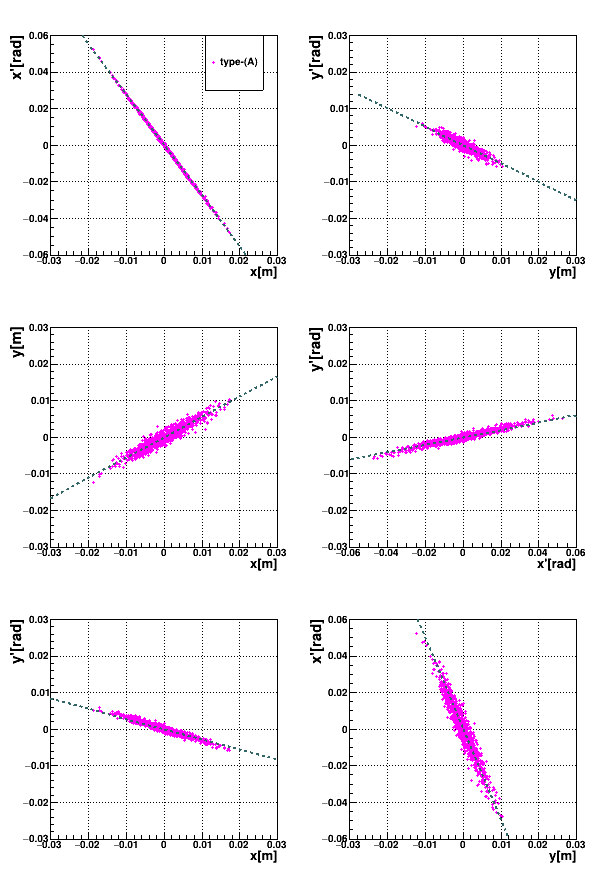}
	 \caption{Strongly \textit{x-y} coupled phase-space generated from Twiss parameters type-(A) in Table~\ref{tab:twiss}. Note that the the method we calculate the Twiss parameters is summarized in our previous study~\cite{Iinuma:2016zfu}.
	 }
      \label{fig:risouPhase}
 \end{figure}

Figure~\ref{fig:risouPhase} shows the phase-space distribution obtained by incorporating finite emittances
$\epsilon_x$ and $\epsilon_y$ in Table~\ref{tab:SITE_beam}~\cite{matsu_master} together with the Twiss parameters of Type-(A) listed in Table~\ref{tab:twiss}.
The straight correlation lines of the ribbon-shaped beam shown in fig.~\ref{fig:ribbon_soukan} are also overlaid. By comparing these, it is confirmed that the beam with finite emittance is still tuned into a flat (ribbon-like) distribution.

\begin{figure}[htb]
 \centering
	 \includegraphics[width=\linewidth]{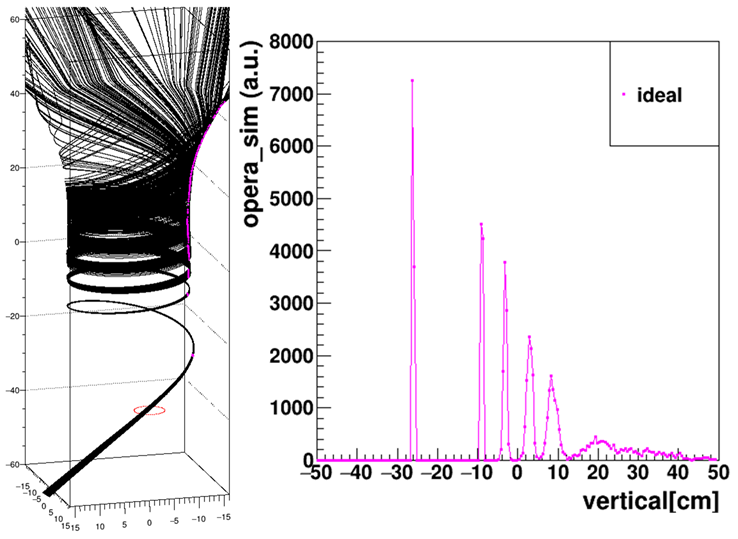}
	 \caption{Left: 3-D beam injection by use of phase space in Fig.~\ref{fig:risouPhase}.
	  Right:Cutout a certain orbital angle to check beam distribution in vertical component. 
	  This is image of wire scan data discussing later.
	 }
      \label{fig:PhaseInj}
 \end{figure}

Figure~\ref{fig:PhaseInj} presents the orbit-tracking results for 1000 particles generated from this phase-space distribution shown in fig.~\ref{fig:risouPhase}.
In the left panel, the pink-colored segment along the 3-D spiral trajectory indicates the region within a limited azimuthal angle range,
$\Delta \phi$=0.02 rad. 
The right panel shows one-dimensional histograms of the beam distribution in the vertical direction.
From these results, the beam distributions corresponding to approximately six turns can be examined at
vertical=-0.3,~-0.1,~-0.02,~0.02,~0.1,~and~0.2 m

\subsection{Possible Twiss Parameter Sets Consistent with the Required \textit{X-Y} Coupling}\label{sec:r-z-theta}

This section examines the permissible ranges of the Twiss parameters and shows that \textit{x-y} coupling is not determined by individual parameter tolerances, but is instead governed by the correlated structure among the eight Twiss parameters. We identify the multivariate relationships that dominate the coupling strength and define the practical acceptance domain for beam injection.

 \begin{figure}[htb]
 \centering
	 \includegraphics[width=\linewidth]{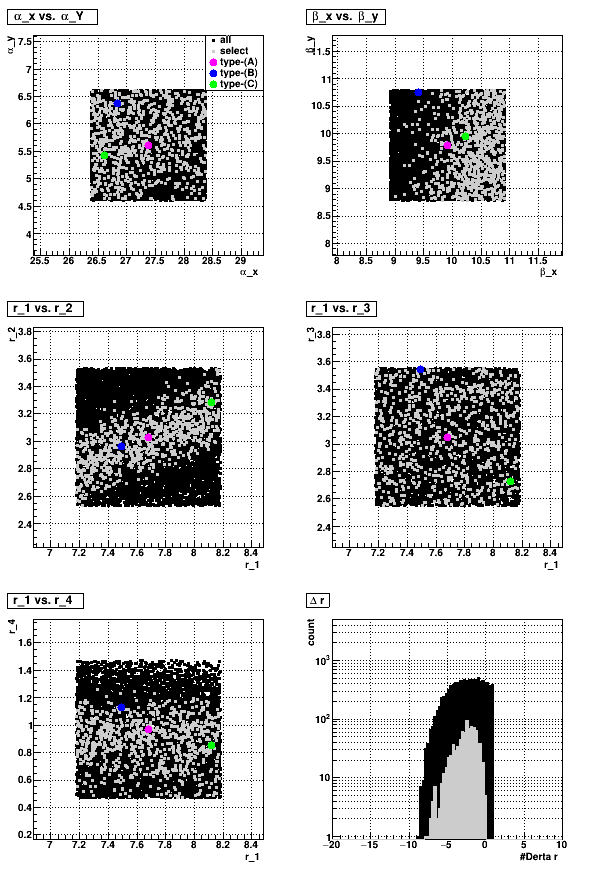}
	 \caption{Black:area of Twiss parameters in Table~\ref{tab:twiss}. 
	 Gray:Accepted area defined in Fig.~\ref{fig:soukanRand}.
	 }
      \label{fig:twissRand}
 \end{figure}

 \begin{figure}[htb]
 \centering
	 \includegraphics[width=\linewidth]{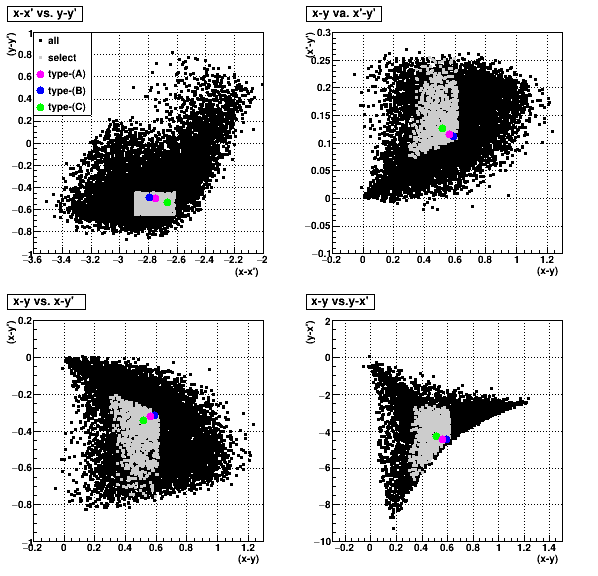}
	 \caption{Black:area of $x$-$x'$,$y$-$y'$, $x$-$y$ and $x'$-$y'$ correlations in Table~\ref{tab:twiss}. 
	 Gray:Accepted area defined in the text.
	 }
      \label{fig:soukanRand}
 \end{figure}
 
Figure~\ref{fig:twissRand} shows the distributions obtained by randomly sampling around a certain range from the central values and ranges listed in Table~\ref{tab:twiss}. 
The distributions of~$\alpha_x$-$\alpha_y$,~$\beta_x$-$\beta_x$,$r_1$-$r_2$,~$r_1$-$r_3$, $r_1$-$r_4$ and $\Delta r\equiv$ ($r_1r_4$-$r_2r_3$)
 are presented under the condition $\Delta r<0$.

\begin{figure}[htb]
 \centering
	 \includegraphics[width=0.8\linewidth]{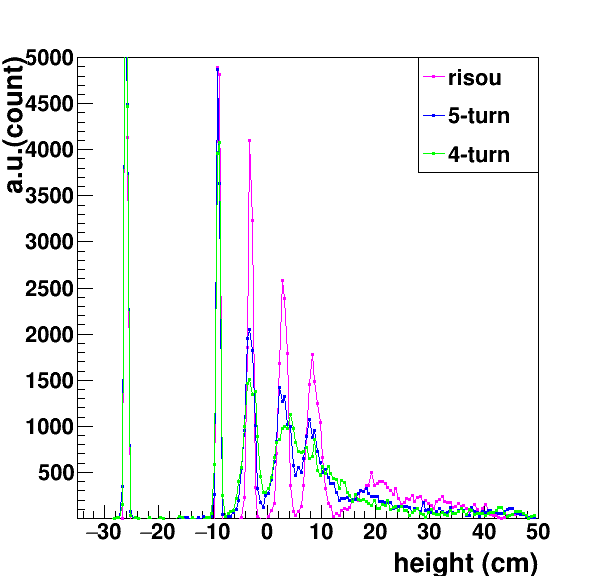}
	 \caption{Examples in case of 5-turns or 4-turns due to different r-z-theta correlations.  
	 }
      \label{fig:SimWire_risou_5107-113-516}
 \end{figure}

\begin{figure}[htb]
 \centering
     \includegraphics[width=0.8\linewidth]{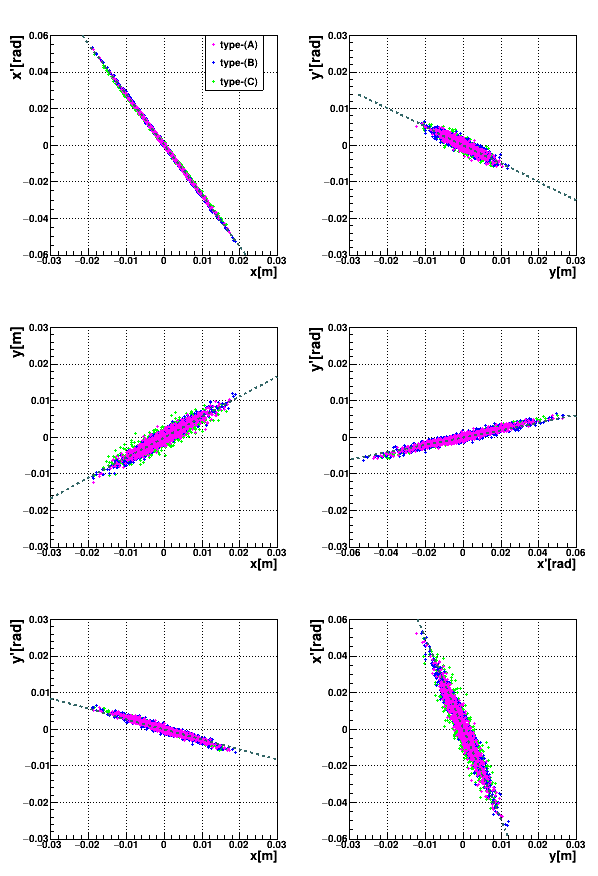}
	 \caption{
     Four–dimensional phase–space ($x,x',y,y'$) at the injection point for several representative Twiss-parameter sets. Although each set yields different injection performance, the clusters overlap substantially in this conventional representation, and the essential differences cannot be clearly identified. This illustrates that the decisive features governing successful spiral injection are embedded in higher-dimensional correlations rather than in individual Twiss parameters alone.}
      \label{fig:risou_5107_516_Phase}
 \end{figure}

For each set of Twiss parameters, a thousand particles' phase space calculation is performed, and the resulting \textit{x-y} correlation are extracted. The distributions of $x$-$x'$, $y$-$y'$,~$x$-$y$, and $x'$-$y'$ are shown in fig.~\ref{fig:soukanRand}.
The black distributions correspond to the sampled Twiss-parameter sets, while the gray regions indicate the acceptable ranges defined
The central pink point represents the reference correlation used in type-(A) of Table~\ref{tab:twiss} and is reflected in fig.~\ref{fig:risouPhase} and fig.~\ref{fig:PhaseInj}.
Similarly, the nearby green and blue points represent the reference correlations used in type-(B) and type-(C) of Table~\ref{tab:twiss}, and are reflected in fig.~\ref{fig:SimWire_risou_5107-113-516} and fig.~\ref{fig:risou_5107_516_Phase}.

By comparing the extent of the gray distributions relative to the black distributions in Figs.~\ref{fig:twissRand} and \ref{fig:soukanRand}, it becomes clear that selecting appropriate combinations of Twiss parameters is more important than the size of the allowed range of each individual parameter.
Accordingly, in the design of a practical transport line, the Twiss-parameter ranges themselves should not be used as primary design metrics. Instead, the design should ensure sufficient coverage of the required correlation space defined by 
$x$-$x'$,~$y$-$y'$,~$x$-$y$,~$x'$-$y'$,~$x$-$y'$, and $y$-$x'$.

The Twiss parameters for type-(B) and type-(C) listed in Table~\ref{tab:twiss} lie far from the ideal point in the Twiss-function space shown in Fig.~\ref{fig:twissRand}.
However, in the \textit{x-y} correlation plane of Fig.~\ref{fig:soukanRand}, both type-(B) and type-(C) move relatively closer to the ideal \textit{x-y} correlation compared with type-(A).
This indicates that the overall multivariate correlation among the Twiss functions is more essential than the individual values of each parameter.
Furthermore, the deviation from the ideal \textit{x-y} correlation directly leads to quantitative changes in the number of turns achieved in the storage-chamber region, as shown in fig.~\ref{fig:SimWire_risou_5107-113-516}, type-(B) and (C).


\begin{figure}[htb]
 \centering
	 \includegraphics[width=0.8\linewidth]{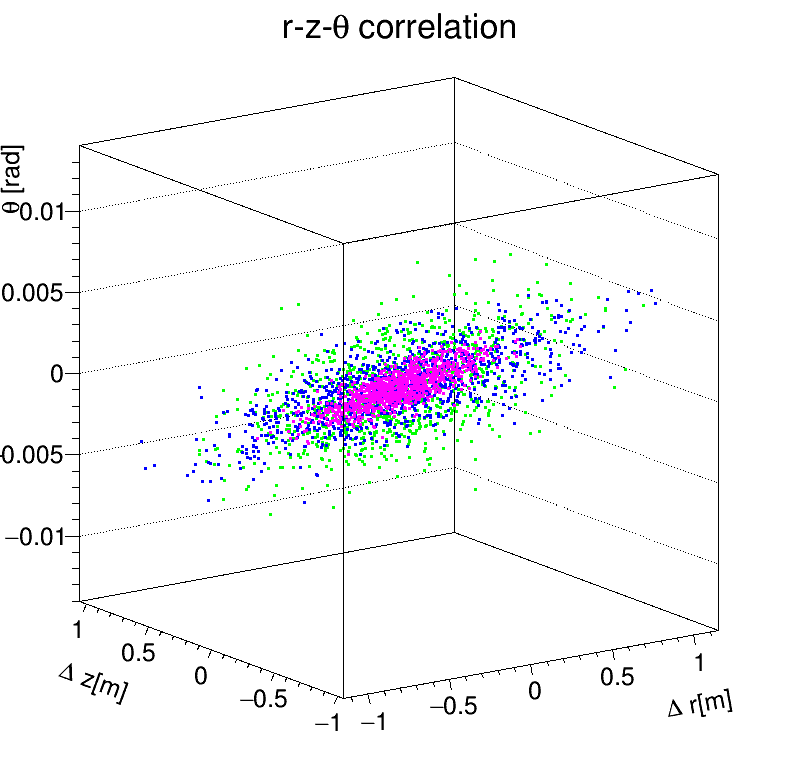}
	 \caption{$R$–$vertical$–$\theta$ representation of the same injection cases as in Fig.~\ref{fig:risou_5107_516_Phase}.~In this coordinate system, which is directly tied to the solenoidal spiral-orbit geometry, distinct differences among the injection conditions become visible.~This clear separation suggests the existence of a dominant feature direction in the multivariate phase space that governs whether the beam reaches the storage region, motivating the use of a data-driven extraction method such as tSVD.}
      \label{fig:r-z-thetaSim}
 \end{figure}

However, but interestingly, comparing the phase-space plots in the beam coordinate system in figs.~\ref{fig:risou_5107_516_Phase}, 
the differences between the three cases are not readily visible. 
Since this representation is not suitable for evaluating the numerical differences summarized in Table~\ref{tab:twiss}, 
we devised an alternative visualization method, as shown in fig.~\ref{fig:r-z-thetaSim}.

Here, $\vec{r}$, the velocity $\vec{v}$, and the corresponding angle $\theta_g$ as in Equation~\ref{eq:eq3} in the laboratory (global) coordinate system as introduced earlier.
We then define the deviations from the reference (central) trajectory as 
$\Delta~r$, $\Delta~vertical$, and~$\Delta~\theta$. 
In this representation, the differences between the three cases become much clearer than in Figs.~\ref{fig:risou_5107_516_Phase}, and
reveals distinct differences in the injection state that directly determine whether the beam can reach the solenoidal storage region.
This observation suggests that there exists a "feature direction"
in the multivariate phase-space that governs the injection outcome.

\begin{figure}[htb]
 \centering
	 \includegraphics[width=\linewidth]{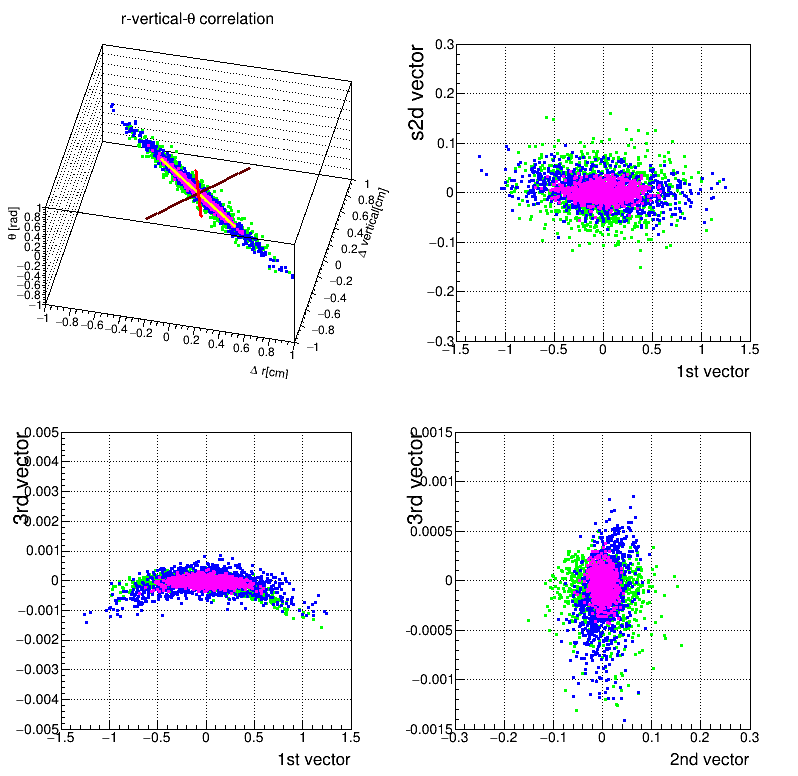}
	 \caption{Left top: $R$–vertical–$\theta$ representation for the same injection cases as in Fig.~\ref{fig:r-z-thetaSim}, using the type-(A) three-axis eigenvectors listed 
This figure shows the projections onto the planes spanned by the eigenvector axes.~The upper-right panel displays the projection onto the plane defined by the first and second eigenvectors.
The lower-left panel shows the projection onto the plane defined by the first and third eigenvectors.~The lower-right panel presents the projection onto the plane defined by the second and third eigenvectors.
}
      \label{fig:r-z-thetaSim_risou_5107_516}
 \end{figure}

To extract this feature direction in a general, data-driven manner,
we introduce the truncated singular–value decomposition (tSVD).
This method has been introduced in Ref.~\cite{Iinuma:ipac2025-wepm029}. 
From the resulting three-dimensional distribution, we extract singular vectors (via SVD) and use them to quantitatively evaluate characteristic features of the phase space.
Figure~\ref{fig:r-z-thetaSim_risou_5107_516} illustrates a representative case in which the ratios of the second and third eigenvalues to the first eigenvalue are on the order of less than 1$\%$;type-(B) and less than 10$\%$;type-(C), respectively. As these ratios decrease—that is, as the first eigenvalue becomes increasingly dominant—the achievable number of turns increases. A comparison of the eigenvalues and the corresponding eigenvectors is summarized~in Table~\ref{tab:eigenTab} and ~\ref{tab:eigenTab2}.

\begin{table}[hbt]
\centering
\caption{Eigen values}
 \begin{tabular}{llll}
\toprule
	item &  type-(A) & (B) & (C)  \\
       \midrule
    1st   & 2.73 & 5.52 & 4.96 \\
    2nd   & 2.04$\times10^{-1}$ & 3.78$\times10^{-1}$ & 6.15$\times10^{-1}$ \\
    3rd   & 1.63$\times10^{-3}$ & 4.83$\times10^{-3}$ & 2.96$\times10^{-3}$ \\
     \bottomrule
\end{tabular}
\label{tab:eigenTab}
\end{table}

Table~\ref{tab:eigenTab2} summarizes the eigen-vector components for Type-(A).~The three-axis vectors for Type-(B) and Type-(C) agree with the Type-A axes within 0.1°.
The full numerical values of all vectors are provided in the Appendix and are omitted here for brevity.
We adopt the ($\Delta r$,~$\Delta\mathrm{vertical}$, $\Delta\theta$) relative to the reference trajectory representation because it provides the most physically meaningful coordinates for describing the magnetic field actually sampled by the injected beam.
$r$ as the absolute value of position vector $\vec{r}$ as in Equation~\ref{eq:eq3} directly characterizes the axisymmetric magnetic-field vector and therefore determines the principal field components relevant to the <$B_{R}L$> correlation and the underlying field index.
Although $r$ and $\mathrm{vertical}$ are not strictly independent, the vertical coordinate specifies the effective field height experienced by the beam and is essential for quantifying the vertical field gradient and the associated \textit{x-y} coupling.
The variable $\theta$ denotes the angle between the position and momentum vectors, and thus serves as a natural parameter for expressing the phase dependence of the injection mismatch and the resulting field–orbit geometry.

This choice is further reinforced by the tSVD decomposition of the ($r$,~$\mathrm{vertical}$, $\theta$) distribution shown in Fig.~\ref{fig:r-z-thetaSim}
The third singular vector becomes fully aligned with the $\theta$ axis, demonstrating that the r–z operating region and the $\theta$ degree of freedom span mutually orthogonal directions.
Accordingly, the phase-space structure can be evaluated in a coordinate system where the in-plane ($r$–$\mathrm{vertical}$) motion and the angular degree of freedom $\theta$ are cleanly separated.

\begin{table}[hbt]
\centering
\caption{Components of eigen vector of type-(A)}
 \begin{tabular}{llll}
\toprule
	item &  1st & 2nd & 3rd  \\
       \midrule
	  $\Delta$~r & 7.49e-01 &6.62e-01 &-3.44e-02  \\
    $\Delta$~$\mathrm{vertical}$  & -6.63$\times10^{-1}$ & 7.48$\times10^{-1}$ & -3.12$\times10^{-2}$\\
    $\Delta~\theta$   &5.12$\times10^{-3}$ & 4.62$\times10^{-2}$ & 9.99$\times10^{-1}$ \\
	\bottomrule
\end{tabular}
\label{tab:eigenTab2}
\end{table}
In this section, we first illustrated the required \textit{x-y} correlation using backward-tracked trajectories of a ribbon-shaped beam, as shown in Fig.~\ref{fig:ribbon_soukan}. 
We then described the procedure for determining the corresponding Twiss parameters that incorporate a finite emittance, and for computing the six correlation plots in the beam-coordinate system, shown in Fig.~\ref{fig:risouPhase}. 
Subsequently, orbit simulations were performed to visualize the resulting beam distributions inside the solenoid magnet, as presented in Fig.~\ref{fig:PhaseInj}. 

We introduced type-(A), representing the ideal case, together with type-(B) and type-(C), which involve small deviations in the phase-space parameters. 
The resulting variation in the number of stored turns was shown in Fig.~\ref{fig:SimWire_risou_5107-113-516}. 
To clearly highlight the differences among type-(A), (B), and (C) in terms of the phase-space distribution at the injection point, we presented the $R$–vertical–$\theta$ correlation plots in figs.~\ref{fig:r-z-thetaSim} and \ref{fig:r-z-thetaSim_risou_5107_516}. 
Using tSVD, we demonstrated that the magnitude of the singular values is directly related to the achievable number of stored turns in the storage region.

In addition to these main discussions, we also showed that the generation of an appropriate \textit{x-y} correlated phase space is governed not by the control of individual Twiss parameters, but by the control of their combinations.

In realistic beamline design, various practical constraints may prevent all of these correlation requirements from being satisfied simultaneously. 
In such cases, it is necessary to define a clear priority order among the correlations and adopt a policy to realize the best achievable 
(or most feasible) phase space—i.e., a situation-dependent “better” solution—even if the optimal one cannot be attained.

This point will be further discussed in Sec.~\ref{sec:x-ydiscussion}, where the phase-space coverage provided by the transport line is examined in more detail.

\section{Principle Demonstration of Three-Dimensional Spiral Beam Injection}
\label{sec:SITE}
Since 2014, we have carried out a principle demonstration experiment of the three-dimensional spiral beam injection, referred to as the Spiral Injection Test Experiment, and have achieved the following milestones in their thesis and , and proceedings contributions:

\begin{enumerate}
\item{Setup and diagnostics of a straight beam line:~\cite{RehmanThesis} and ~\cite{RehmanLINAC18,RehmanIQBS2019,matsushita:ipac2021-mopab256,Matsushita:ipac2023-mopa118,Iinuma:IPAC2018-TUPML060,Iinuma:ipac2025-thpm099}}
\item{Control strongly X–Y-coupled phase space using a rotating quadrupole magnet:~\cite{RehmanThesis} and ~\cite{RehmanLINAC18,RehmanIQBS2019,matsushita:ipac2021-mopab256,Matsushita:ipac2023-mopa118,Iinuma:IPAC2018-TUPML060,Iinuma:ipac2025-thpm099}}
\item{Visualization and image-based identification of spiral beam orbits, and setup quantitative beam distribution measurement (wire-scan, etc.):~\cite{RehmanThesis},~\cite{matsu_master},~\cite{rehman:ipac2021-mopab162}}
\item{Dedicated data  for beam cross-sections with several types of \textit{x-y} coupled conditions:This manuscript,}
\item{Demonstration of beam storage using a vertical kicker and weak focusing fields; the final goal of this experiment. This is beyond the scope of this paper, but result is available see~\cite{matsuPRL}.}
\end{enumerate}

Table~\ref{tab:Tab1} summarizes the basic parameters of this experiment.
\begin{table}[hbt]
\centering
\caption{Basic parameters of this experiment}
\label{tab:Tab1}
\begin{tabular}{lll}
\toprule
Item & DC mode & Pulse mode \\
\midrule
Main magnetic field & \multicolumn{2}{c}{8~mT} \\
Weak focusing $N$ & \multicolumn{2}{c}{$(1.0\sim5.6)\times10^{-2}$} \\
\midrule
Beam & \multicolumn{2}{c}{$e^{-}$, 80~keV} \\
 & DC & 100~ns pulse~\cite{matsuPRL} \\
Turn period & \multicolumn{2}{c}{5.0~ns} \\
Ring radius & \multicolumn{2}{c}{0.11~m} \\
\midrule
Rotating quads & \multicolumn{2}{c}{3} \\
\midrule
Kicker~\cite{matsuPRL} & NA &  \\
Duration time & NA & 140~ns \\
Peak current & NA & 45~A \\
\bottomrule
\end{tabular}
\end{table}

\begin{figure}[htb]
 \centering
	 \includegraphics[width=\linewidth]{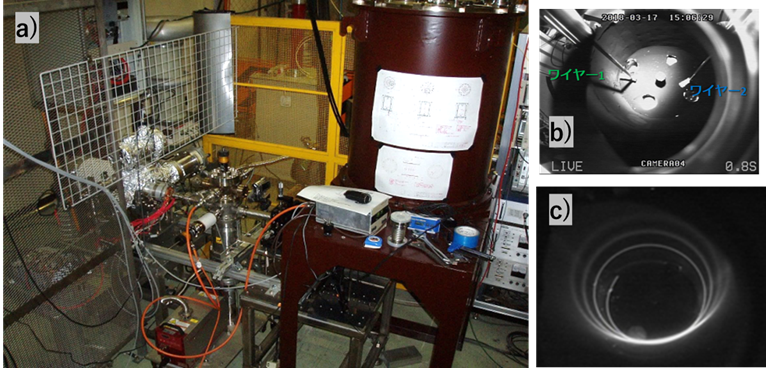}
	 \caption{(a) Overall view of the experimental apparatus.
(b) View of the vacuum chamber inside the storage solenoid magnet, photographed from an upper viewport using a wide-angle lens camera. The two rods visible in the image are feedthroughs for the wire scanner.
(c) Image of the three-dimensional spiral trajectory visualized via ionization light emission in nitrogen gas inside the storage chamber. This image was taken before installation of the kicker system, allowing the entire trajectory to be observed.
	 }
      \label{fig:SITEAll}
 \end{figure}

%
Figure~\ref{fig:SITEAll}(a) shows an overview photograph of the beamline. 
Figure~\ref{fig:SITEAll}(b) shows the inside of the storage chamber, together with two wire scanners (for DC-beam operation) 
used to measure the beam distribution along the solenoid axis.

A chamber monitor is installed to visualize the beam trajectory inside the vacuum chamber of the storage magnet. 
Figure~\ref{fig:SITEAll}(c) shows an image of the ionization light emitted along the trajectory of a DC electron beam in nitrogen gas, 
recorded with a high-sensitivity camera. As discussed in the previous section, 
injecting a beam with appropriately tuned \textit{x-y} coupling enables clear visualization of a spiral trajectory inside the storage chamber.

The \textit{x-y} coupling was tuned online while monitoring the coupled beam using three rotating quadrupole magnets shown in Figs.~\ref{fig:BeamLinePict}, 
together with beam-profile monitors installed at the end of the straight section (Straight-Monitor) and upstream of the storage-magnet injection point (Bend-Monitor).

The achievable range of \textit{x-y} coupling in this beamline is discussed in Sec.~\ref{sec:rotQLine}, and the post-injection beam measurement results are presented in Sec.~\ref{sec:rotQLine}.

\subsection{From the Electron Gun to the Straight-Section Beam Diagnostics}

In this section, we discuss the results obtained from DC electron-beam operation carried out between 2020 and 2023, 
including the tuning of \textit{x-y} coupling and the measurements of three-dimensional spiral trajectories inside the storage magnet. 
The trajectories were observed using a high-sensitivity camera installed above the storage chamber.
It is worth noting that it took about one year to transport the 80~keV electron beam to the most downstream monitor in the straight section, 
and approximately four years from the start of injection operation into the storage chamber in 2016 to reach the stage of measuring an \textit{x-y} coupled beam.

\begin{figure}[htb]
 \centering
	 \includegraphics[width=\linewidth]{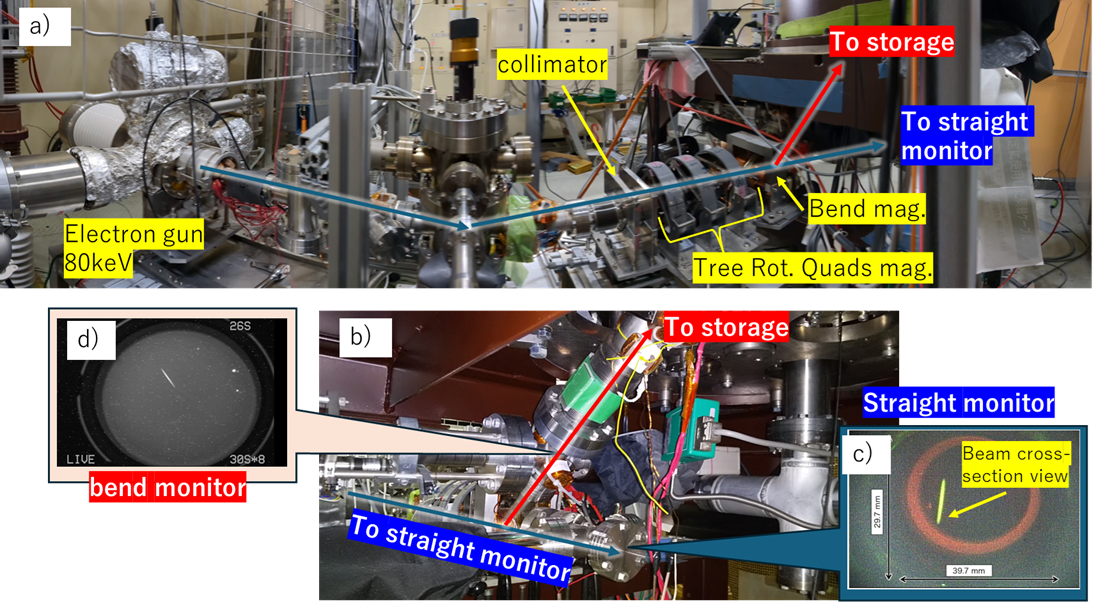}
	 \caption{a) Entire view of straight section. b) view from the straigh end. 
	    c) Pictures of straight-monitor and bend-monitor. 
	     Measurements of beam cross-section at the Straight-monitor are utilized for \textit{x-y} coupling adjustment.    
	 }
      \label{fig:BeamLinePict}
 \end{figure}

Figure~\ref{fig:BeamLinePict}(a) shows a photograph covering, in a single field of view, the beamline from the upstream electron gun to the most downstream beam diagnostic device in the straight section.

The collimator located at the center serves both to scrape beam halo from upstream and to maintain differential pumping. 
While the electron-gun side is kept at a pressure of $1\times 10^{-6}$~torr, the storage chamber is filled with nitrogen gas at approximately
$7\times10^{-3}$~torr. 
Ionization light emitted along the beam trajectory is recorded with a high-sensitivity camera, enabling visualization of the three-dimensional spiral injection.

If the distribution of individual charged particles at the collimator is expressed in the beam coordinate system as
$\vec{X}_{in}=(x_{in},x'_{in},y_{in},y_{in})^{t}$) 
and the distribution at the screen located at the downstream end of the straight section is expressed as
$\vec{X}_E=(x_E,x'_E,y_E,y'_E)^{t}$), then

\vspace{-2mm}
\begin{equation}
	\vec{X_e}=M_{i}\vec{X_{in}}
\label{eq:mat_straight1}
\end{equation}
and is related to the matrix elements of the transfer matrix
$M_i$~for a given setting ($i^{th}$) through the following relation:

\vspace{-2mm}
\begin{equation}
	\begin{bmatrix}
		 x ^{i} \\ \nonumber
		 x'^{i} \\ \nonumber
		 y ^{i} \\ \nonumber
		 y'^{i}
	\end{bmatrix}_{E}  =
            \begin{bmatrix}
		    M_{11}   & M_{12}  & M_{13}   & M_{14}  \\
		    M_{21}   & M_{22}  & M_{23}   & M_{24}  \\
	            M_{31}   & M_{32}  & M_{33}   & M_{34}  \\
		    M_{41}   & M_{42}  & M_{43}   & M_{44}  \\
	    \end{bmatrix} 
	\begin{bmatrix}
		 x \\ \nonumber
		 x'\\ \nonumber
		 y \\ \nonumber
		 y'
	\end{bmatrix}_{in}  
\label{eq:mat_straight2}
\end{equation}
Using the $\Sigma$ matrix formulation commonly adopted in beam dynamics,

\vspace{-2mm}
\begin{equation}
	\Sigma_E=M_i\Sigma_{in}~M_i^{t}
\label{eq:eq3}
\end{equation}

When extended to $n$~particles,
\vspace{-2mm}
\begin{equation}
	\vec{X}_{in}^{t}\Sigma_{in}^{-1}~\vec{X}_{in}=1
\label{eq:eq4}
\end{equation}
The projected moments ⟨$xx$⟩, ⟨$yy$⟩, and ⟨$xy$⟩ measured at the straight-section monitor can be written as linear combinations of the these ten independent elements.
The explicit expressions are provided in Appendix.

\subsection{Beam Monitor}\label{sec:Monitor}
The beam from the electron-gun is transported through a sequence of the straight section and bending elements before entering the solenoid magnet leading to the storage volume.
In this experiment, only two beam profile monitors are available at the end of the straight section and at the injection point down stream of the bending magnet as shown in fig.~\ref{fig:BeamLinePict}.

A profile monitor located in the straight section provides two-dimensional beam images for reconstructing the projected second moments ⟨$xx$⟩, ⟨$yy$⟩, and ⟨$xy$⟩.
These data serve as the experimental inputs for evaluating the \textit{x-y} coupling and for comparing with the simulated beam distributions in the beam coordinate.

The monitor images exhibit a characteristic tilt and aspect-ratio change when the incoming beam contains a non-negligible \textit{x-y} coupling component.
This behavior should be confirmed by calculation by use of Equations~\ref{eq:mat_straight1},~\ref{eq:mat_straight2}. The consistency between the measured profiles and the calculated moments provides a direct validation that our transfer-matrix description of the beamline is accurate.

The monitor consists of a scintillating screen and a camera system with an effective spatial resolution of 0.1 mm, sufficient for resolving the projected beam ellipticity and its rotation.

\begin{figure}[htb]
 \centering
	 \includegraphics[width=\linewidth]{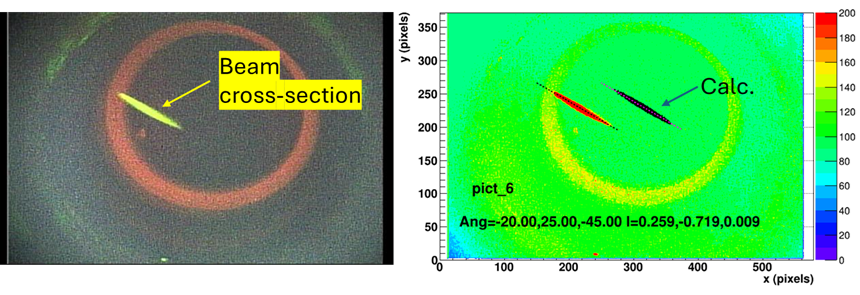}
	 \caption{Left:Sample view of Straight-Monitor. 
	 Analysis of crossection view as well as comparison of simulation results.
	 }
      \label{fig:Straight-monitor}
 \end{figure}

Figure~\ref{fig:Straight-monitor} shows an image acquired by the straight-section end monitor, together with the corresponding pixel-intensity histogram derived from the image data.
The black distribution in the figure represents the simulation results discussed later.

\begin{figure}[htb]
 \centering
	 \includegraphics[width=\linewidth]{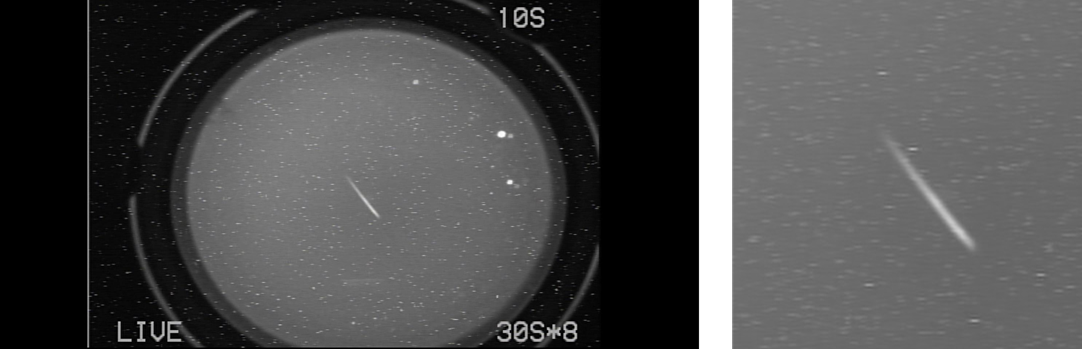}
	 \caption{Left: Original view at the bend monitor. Right: Averaged zoom-up view to analyze.
	 }
      \label{fig:BendView0}
 \end{figure}

The orbit was then bent upward by 45° using a bending magnet, and the beam profiles at the injection point into the storage magnet (Bend monitor) are shown in fig.~\ref{fig:BendView0}.
While the transverse beam images at the straight-section monitor could be adjusted to fit well within the camera pixel area, the images at the Bend monitor appear small in the pixel area and have weak signal due to limitations of the camera performance. 
Therefore, the \textit{x-y} coupling was primarily tuned using the straight-section beam monitor.

Figure ~\ref{fig:DC1_sim} compares the measured beam profiles at the straight-section monitor and the injection point with simulations based on the reconstructed $\sigma$ matrix and transport-line model.

 The Straight-monitor image shown in Fig.~\ref{fig:BeamLinePict}(c) presents the beam transverse profile at the downstream end of the straight section.
From a single image, three quantities,
$\langle xx \rangle$,$\langle yy \rangle$, and $\langle xy \rangle$, can be measured.

The explicit form of $T_{mom}$
 is given in the Appendix and in Ref.~\cite{Xiao}.

The $\sigma$ matrix at the upstream collimator in the straight section can thus be obtained.
Since the $\sigma$~matrix describing the beam phase space consists of ten independent variables, 
in principle at least ten sets of image data are sufficient.

\subsection{Generation and Diagnostics of Strongly X–Y Coupled Beams}\label{sec:rotQLine}

Using the three rotating quadrupole magnets installed in the straight section, 
a Q-scan was performed with the magnets set as normal quadrupoles (
$\phi_Q$=0).
From this scan, the emittance at the beam starting point (collimator section), the corresponding Twiss parameters, and the coupling coefficient
$t_{cup}$~were determined.
The results are summarized in the following table.
\begin{table}[!hbt]
   \centering
   \caption{Twiss parameters at collimator}
   \begin{tabular}{lccc}
       \toprule
	   symbol     & x              & y                      & dimension \\
      \midrule
	   $\alpha$   & -3.2~$\pm$~0.8 & -2.8 $\pm$ 0.3         &  rad\\ 
	   $\beta$    & 1.8~$\pm$~0.5  & 1.6 $\pm$ 0.5          & m\\ 
	   $\gamma$   & 7.0~$\pm$~1.0  &  6.4$\pm$0.4           &rad/m \\ 
	   $\epsilon$ & 2.8~$\pm$~0.2  &  3.0$\pm$0.5 &1$^{-7}$ rad.m     \\ 
	   $ t_{cup}$\cite{Xiao}    & $<$~0.01 & $<$~0.01    \\ 
       \bottomrule
   \end{tabular}
   \label{tab:SITE_beam}
\end{table}

To evaluate the validity of the \textit{x-y} coupling control, simulations of the beam transverse profile at Monitor-1 were performed using 
the reconstructed $\sigma$~matrix.
The rotation angles and strengths ($k$ values) of the rotating quadrupoles were then varied to generate 21 different 
transport-matrix settings (with an additional 40 cases shown in the Appendix), 
and the resulting \textit{x-y} images at the downstream end of the straight section were compared with the simulations.

\begin{figure}[htb]
 \centering
	 \includegraphics[width=0.9\linewidth]{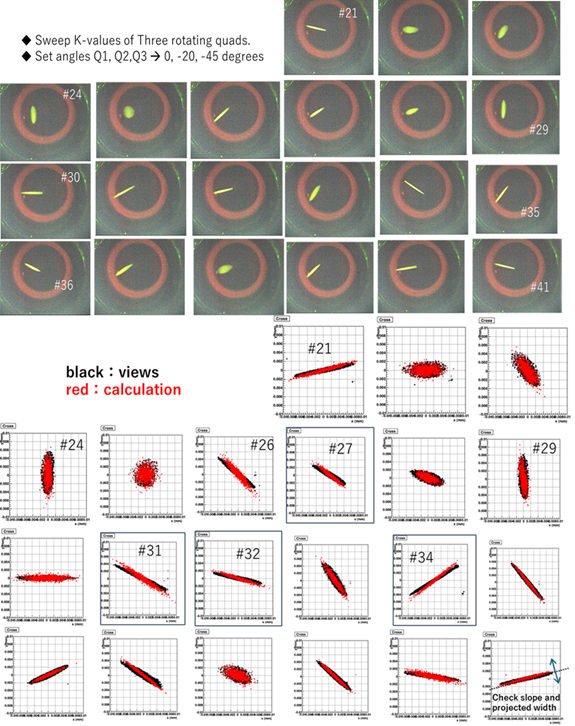}
	 \caption{Up:Beam cross-section views at Monitor-1.
	 Bottom: Comparison of beam cross-sections by changing beam line settings.  The method used to determine the slope and width of the distribution is also illustrated in the lower-right corner.
	 }
      \label{fig:fig13}
 \end{figure}

The upper panels of Fig.~\ref{fig:fig13} show the 21 beam images measured at Monitor-1 at the end of the straight section. During these measurements, the quadrupole rotation angles were fixed at 
0,~-20, and -45 degrees, while the $k$values were scanned. 
From the analysis of Fig.~\ref{fig:fig13}, the relative displacement of the beam centroids at Monitor-1 is within 3~mm~ as shown in Appendix, 
indicating that the axis misalignment (alignment accuracy) of each rotating quadrupole is within 100~$\mu$m.

\begin{figure}[htb]
 \centering
	 \includegraphics[width=6cm]{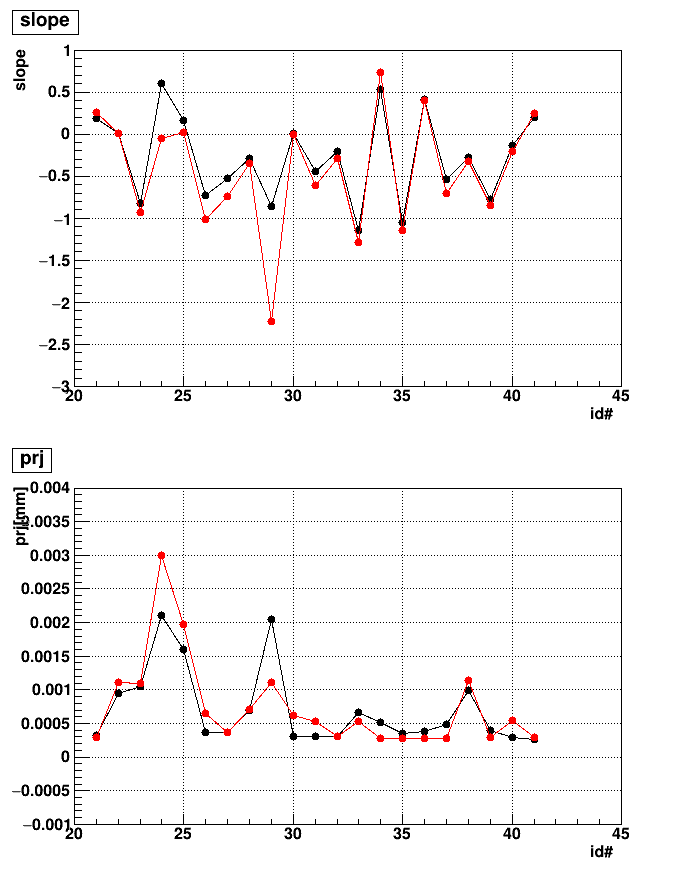}
	 \caption{
     Top: Slope of the distribution shown in Fig.~\ref{fig:fig13}.
Bottom: Width of the distribution, as illustrated by the example shown in the lower-right corner of Fig.~\ref{fig:fig13}.
	 }
      \label{fig:fig14}
 \end{figure}

The results are shown in Fig.~\ref{fig:fig13}, where the black distributions represent the image data 
and the red distributions represent the simulation results.
Figure~\ref{fig:fig14} further shows the slopes of the black and red distributions (i.e., the
\textit{x-y} correlation slopes) and the widths projected along the slope axis. 
Considering the angular setting accuracy of each rotating quadrupole ($\pm$0.5~degrees) and the uncertainty in
$k$ derived from the current control accuracy (0.01 A), together with the configuration of the three consecutive quadrupoles 
and the drift space to Monitor-1, the simulations reproduce the experimental data well (see Appendix).

Thus, using beam transverse images for individual settings obtained with Monitor-1 installed in the straight-section beamline, we demonstrate that the beamline system can be understood and controlled as an integrated whole.
In other words, calculated estimates can be provided for other phase-space parameters that cannot be measured directly.

 \begin{figure}[htb]
 \centering
	 \includegraphics[width=\linewidth]{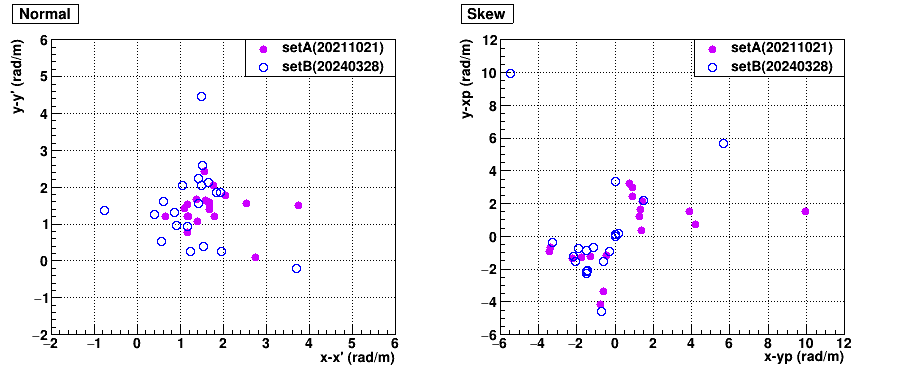}
	 \caption{
     Slopes of the $x$-$x'$ and $y$-$y'$ correlations (left), and those of the $x$-$y'$ and $y$-$x'$ correlations (right), calculated using the data acquisition parameters of Fig.~\ref{fig:fig13}. 
The correlations in the normal transverse planes remain positive and do not simultaneously satisfy convergent conditions, indicating that the beamline behavior deviates from the nominal design expectation. 
In contrast, convergent tendencies appear in the skew correlations, suggesting that the observed beam transmission may be supported by skew-plane focusing effects, which provide a possible explanation for the successful beam storage obtained under realistic conditions.
     }
\label{fig:sokan-50}
\end{figure}
 
As explained earlier, the image quality is not as high as that obtained at the straight-section monitor; nevertheless, it provides a useful consistency check for our understanding of the transfer matrix of the entire injection beamline, including the bending magnet.
Although not shown in the main text, consistency between the bend monitor images near the injection point and the simulation results has also been confirmed; details are provided in the Appendix.

 In the following sections, we discuss the evaluation of three-dimensional spiral trajectories inside the storage chamber using a DC electron beam.
Results of pulsed-beam injection and storage experiments employing the kicker system are reported elsewhere and are beyond the scope of this paper.

\section{Experimental Verification of the Three-Dimensional Spiral Injection}
This section presents the experimental observations that validate the fundamental mechanism of the three-dimensional spiral injection scheme.
The results are organized into three logical layers:
\begin{enumerate}
\item Validation of the injection principle through direct visualization of multi-turn spiral trajectories.
\item Experimental verification of the X–Y coupling design guideline using real beam data.
\item Reconstruction and interpretation of the \textit{x–y} phase space, followed by an evaluation of the $r$–vertical–$\theta$ representation introduced in Sec.~\ref{sec:r-z-theta}.
\end{enumerate}

Together, these results demonstrate that the key elements of the spiral-injection concept—strong X–Y coupling, appropriate phase-space shaping, and geometric overlap with the solenoidal acceptance—are realized experimentally.

\subsection{Visualization of Multi-Turn Spiral Trajectories}

\begin{figure}[htb]
 \centering
 \includegraphics[width=\linewidth]{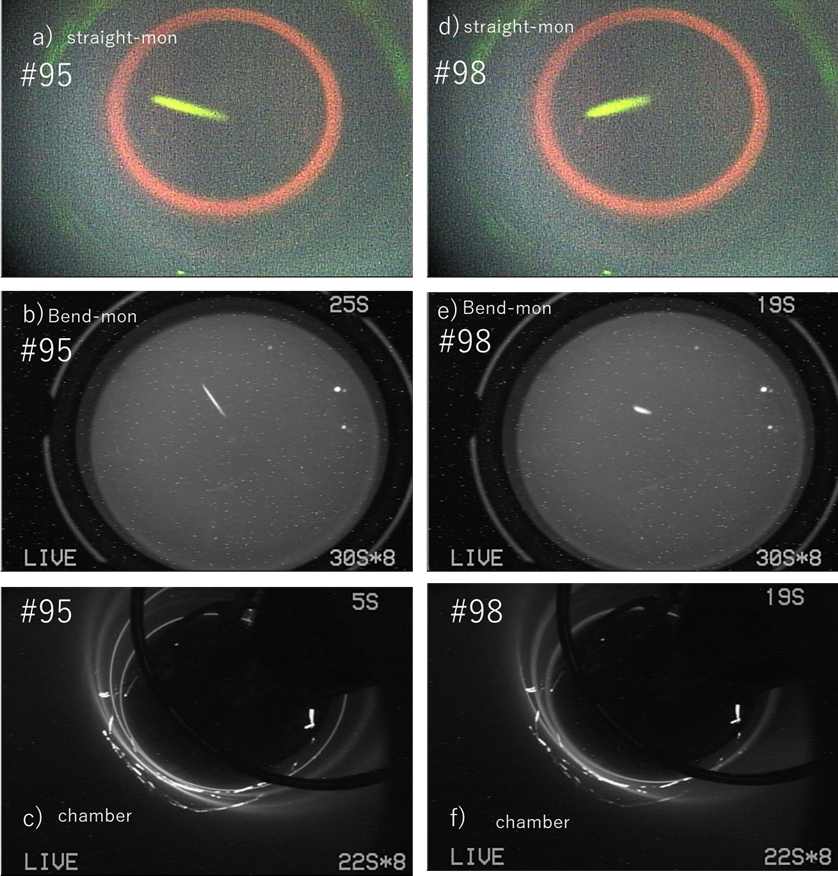}
	 \caption{Comparing the two-dimensional histograms of pixel intensities derived from the CCD images in (a),(d) and (b),(e). (c) and (f) are three-dimensional spiral trajectories' Views of in the storage chambers.}
      \label{fig:DC1_full}
 \end{figure}

Spiral injection was tested by scanning the rotation angles ($\phi_Q$) and strengths ($k_Q$) of the three rotating quadrupoles while monitoring transverse beam profiles at the straight-section monitor and at the injection point.
Table~\ref{tab:setQ123} summarizes representative settings which are discussed in this paper. Three examples (ID 95 and 98) are shown in Fig.~\ref{fig:DC1_full}.
\begin{table}[!hbt]
   \centering
   \caption{
   Lists the rotated-quadrupole settings used for the three data sets (67, 95, and 98) shown in Fig.~\ref{fig:DC1_wire195}.
The 2024 settings correspond to the post-upgrade beamline configuration, while the 2021 values (3,4,6 and 9) are included for comparison.
The associated 3-D spiral injection images are presented in Fig.~\ref{fig:Others}.
   }
   \begin{tabular}{l|c|c|c}
       \toprule
	   ID$\#$& $\psi_Q$(deg.) & $k_Q$ & data/Fig. \\
      \midrule
      67  &  0,~0, 0  &  0.0,~-0.0,~0.0  &2024 Mar. \\ 
        95  &  0,~20,~-45  & -20.3,~-48.3,~7.11 & in \\ 
	   98  &  0,~20,~-45  & -81.2,~38.4,~7.11 & fig.~\ref{fig:DC1_full},~\ref{fig:NoQ}   \\ 
      \midrule
	   3   & 0,~15,~45  & -22.2,~-70.95,~ -22.2 &2021 Feb.  \\ 
	   4   & 0,~15,~45 & -26.29,~-72.98,~0.0  &  in \\ 
	   6   & 0,~15,~45  & -22.2,~-70.95,~ -22.2 & fig.~\ref{fig:Others}\\ 
	   9   & -20,~25,~-45 & -26.29,~-72.98,~0.0   &\\ 
       \bottomrule
   \end{tabular}
   \label{tab:setQ123}
\end{table}


Figures ~\ref{fig:DC1_full}(c) and \ref{fig:DC1_full}(f) show ionization-light images of the beam trajectory inside the storage chamber.
The beam profiles measured at Monitor-1 in the straight section, corresponding to the case where the bending magnet is turned off, are shown in (a) and (d).
While beam injection operation, the beam was bent upward by 45° using a bending magnet. There is another monitor to take the beam profiles at the injection point into the storage magnet (Bend monitor), and they are shown in (b) and (e). 
Comparing panels (a) and (d) with (b) and (e), it is clear that the transverse beam images measured at the straight-section monitor are better suited for tuning the rotating quadrupoles, as the beam profiles are well contained within the camera pixel area.
In contrast, the images at the bend monitor occupy a much smaller pixel area and exhibit weaker signal intensity due to limitations in the camera performance.

\begin{figure}[htb]
 \centering
     \includegraphics[width=\linewidth]{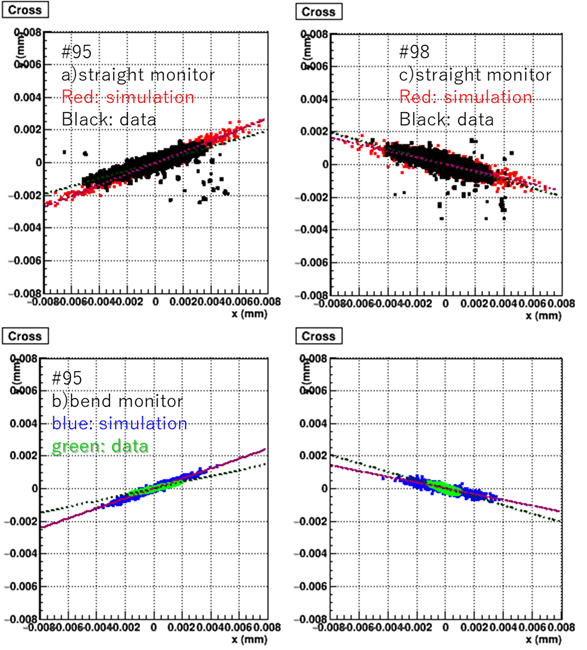}
	 \caption{
     (a) and (c) show cross-sectional views of the beam profile at the straight monitor for cases $\#$95 and $\#$98, respectively. (b) and (d) present the corresponding cross-sectional views at the bend monitor, consistent with the beam monitor images shown in Fig.~\ref{fig:DC1_full}.
	 }
      \label{fig:DC1_sim}
 \end{figure}

Figures~\ref{fig:DC1_sim}(a) and (c) compare the measured beam profiles with simulations based on the reconstructed $\Sigma_E$ matrix using a straight transport-line model.
The black dots represent the measured data shown in Figs.~\ref{fig:DC1_full}(a) and (d), while the red points correspond to the simulations.~Similar comparisons based on the bent transport-line model are presented in Figs.~\ref{fig:DC1_sim}(b) and (d), using the measured data shown in Figs.~\ref{fig:DC1_full}(b) and (e).

Because a clear image to judge the \textit{x-y} coupling was required, we decided to tune the beam using the straight-section beam monitor primarily.
From these measurements, we confirmed a good agreement between the calculated beam profiles and the observed ones.
For beam setting $\#$95, the measured tilt of the \textit{x-y} image matches the design requirement within 5

Other phase-space parameters contributing to the \textit{x–y} correlation cannot be directly measured in this beamline.
However, as discussed in the previous section, the transport matrix 
$M$~of our beamline is well understood.
Therefore, the overall phase-space consistency, as represented by the 
$\Sigma$-matrix, is validated.
This point will be discussed in more detail in the following sections as in $\Sigma$-matrix.

Importantly, as in fig.~\ref{fig:DC1_full}(c), (f), the sign of the \textit{x-y} correlation influences the spiral geometry. 
For ID 95 in fig.~\ref{fig:DC1_full}~(a), the transverse image at Monitor-1 exhibits the expected slope, and the resulting four-turns spiral trajectory is well confined as in (c).
For ID 98, where the slope is reversed~(d), the spiral trajectory becomes visibly distorted down to three-turns even though the straight-section image looks similar in size.

\begin{figure}[htb]
 \centering
 \includegraphics[width=\linewidth]{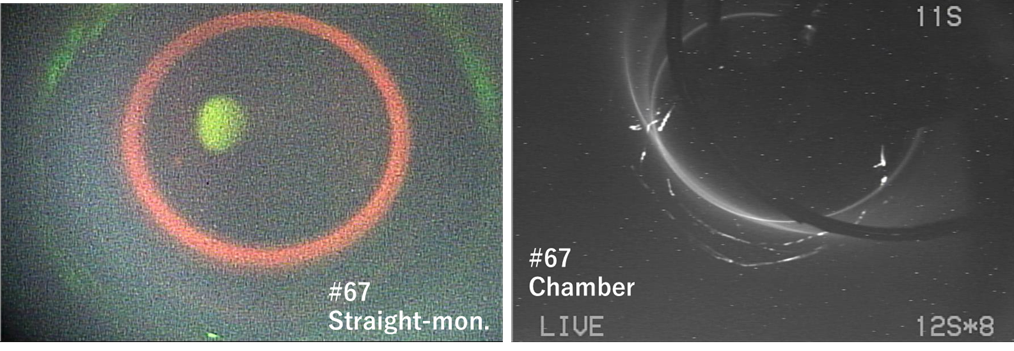}
	 \caption{Example of injection without phase-space shaping.
Left: Beam image at the most downstream straight-section monitor, showing an approximately circular distribution due to the absence of shaping.
Right: Ionization-light image inside the storage chamber.}
      \label{fig:NoQ}
 \end{figure}

Figure~\ref{fig:NoQ} depicts a case of all rotating quadrupoles are turned off, and the electron-gun beam is transported in free space without any \textit{x-y} correlated shaping.
Compared with the shaped cases (IDs 95 and 98), the injected ensemble shows much stronger divergence and fails to form a coherent spiral trajectory.	

\begin{figure}[htb]
 \centering
	 \includegraphics[width=\linewidth]{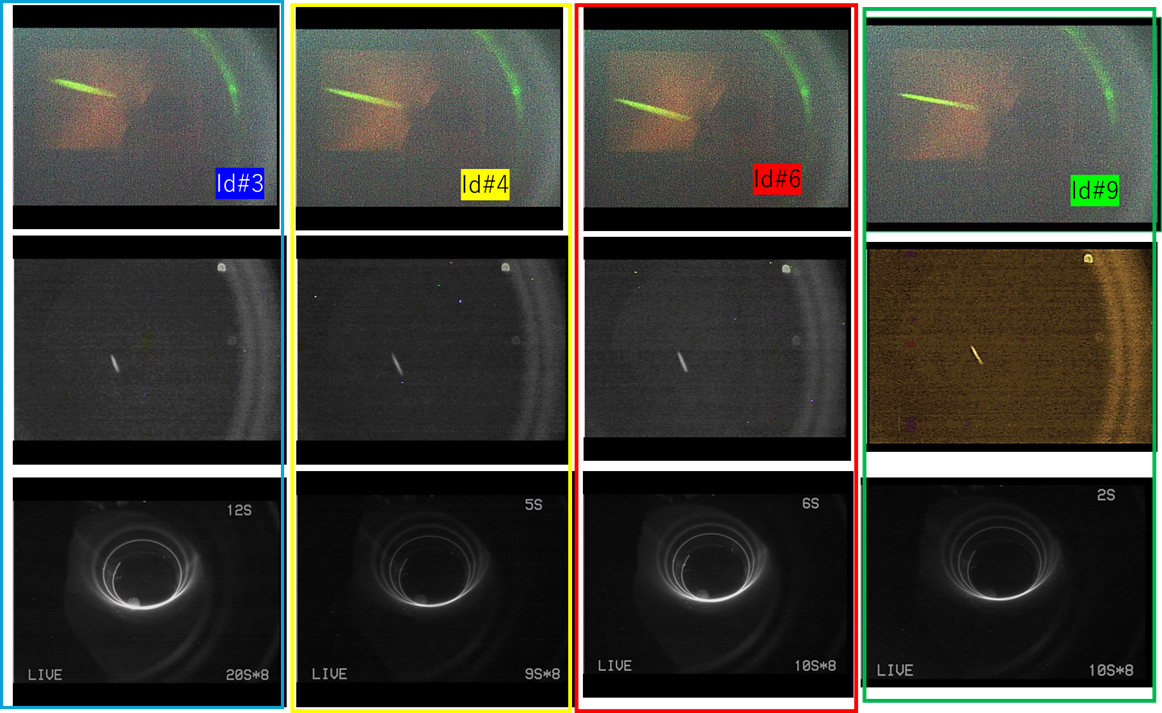}
	 \caption{
     Other examples, corresponding to ID$\#$3, 4, 6, and 9 from the left, were obtained in February 2021.~Note that the images for both the straight-section monitor and the bend monitor differ from those shown in Figs.~\ref{fig:DC1_full} and \ref{fig:NoQ}.
This is because these data were taken after modifications to the beamline in 2021, during which the straight-section monitor and the bend monitor were replaced.
The views of the storage chamber are also different, as a kicker system was installed in the storage chamber in 2021, which changed the camera installation position.
In addition, wire scanners were introduced to enable more quantitative measurements of the beam distribution along the solenoid axis.
	 }
      \label{fig:Others}
 \end{figure}

This qualitative behavior already indicates that the spiral-injection performance is governed by the detailed structure of the multivariate phase-space correlations captured by $\Sigma$-matrix.
For reference, the quadrupole settings used in an earlier attempt in 2021 are also listed for comparison. Their quadrupole parameters and \textit{x-y} slopes are summarized in Table~\ref{tab:setQ123}, and the corresponding images of the three-dimensional spiral injection for each condition are shown in Fig.~\ref{fig:Others}.






Further results for IDs 
$\#3,4,6$,~and $\#9$ listed in Table~\ref{tab:setQ123} are shown in fig.~\ref{fig:Others}
.
This figure presents examples obtained prior to the beamline modifications.
It is included as a reference to illustrate a wider range of beam-profile variations, and therefore is not discussed in depth in the main text.

For IDs 
$\#3,4,6,9$, and $\#95$, the quadrupole rotation angles were set and the strengths 
$k_Q$~were tuned to approach the target \textit{x-y} images defined by the design. 
Interestingly, provided \textit{x-y} coupling close to the design values, with all quadrupole rotation angles changed.

These results indicate that, even with different rotating-quadrupole settings, 
there exist target settings that bring the \textit{x-y} beam profile close to the ideal one, 
and that the corresponding trajectory images inside the storage chamber can be tuned to a state where approximately four-turns are visible.
In other words, multiple Twiss-parameter scenarios can realize the prescribed beam phase space, which is consistent with the discussion earlier.

The results show a clear and systematic trend:
\begin{enumerate}
  \item  Cases with correlation parameters close to the ideal values (e.g., IDs 3, 4, 6, 95)~$\rightarrow$~produce 3–4 turns in the storage region,
\item Cases with significant deviations—especially sign reversals in x–y' and y–x' $\rightarrow$~result in broadened or distorted spiral trajectories, reducing the number of turns. 
\end{enumerate}

This demonstrates experimentally that the four-dimensional correlation structure~($x,x',y,y'$)~not individual Twiss parameters alone—governs the success of the spiral injection.

\subsubsection{Quantitative Confirmation via Wire-Scan Measurements}

While the ionization images qualitatively reveal the multi-turn spiral, a quantitative evaluation of the beam distribution along the solenoid axis is obtained from wire-scan measurements.

 \begin{figure}[htb]
 \centering
  \includegraphics[width=\linewidth]{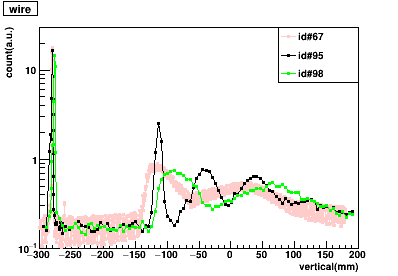}
	 \caption{
     Wire-scan data for settings 95 and 98 were taken together with an additional data set, “67,” in which all quadrupoles were switched off and the beam was injected after free-space transport.
Because the acquisition mode for 67 differs from that of 95 and 98, the three data sets are not plotted on the same manner but still comparable. 
Interestingly, the measured beam profile for setting 98 (reversed \textit{x-y} coupling) closely resembles that of 67.
	 }
      \label{fig:DC1_wire195}
 \end{figure}
Figure~\ref{fig:DC1_wire195}) depicts the wire-scan measurements for settings 95 and 98, an additional data set, “67.” 
Again, $\#$67 was acquired under a condition in which all quadrupoles were turned off and the beam was injected after propagating through free space.
Since the wire-scan acquisition mode for 67 differs from that used for 95 and 98, these data cannot be displayed on the same manner plot, but still comparable. 

\subsection{Model–Measurement Consistency}\label{result2}

\begin{figure}[htb]
 \centering
	 \includegraphics[width=\linewidth]{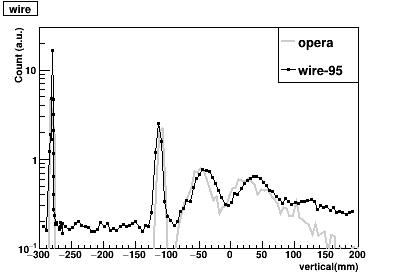}
	 \caption{Comparison with simulation and real data $\#$95 as in Fig.~\ref{fig:DC1_wire195}.
The four turns observed in Fig.~\ref{fig:DC1_full}(c) are also confirmed by the wire-scan measurement and are consistent with the calculation results (labeled OPERA). }
      \label{fig:95_wiresim}
 \end{figure}

\begin{figure}[htb]
 \centering
	 \includegraphics[width=\linewidth]{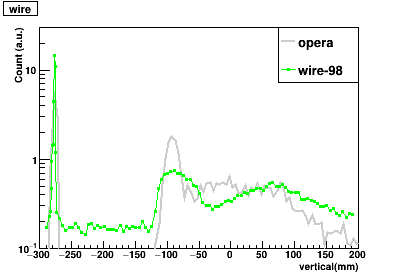}
	 \caption{
     Figure~\ref{fig:DC1_wire195} shows a comparison between the simulation and the measured data for case $\#98$.~A broader vertical distribution, reflecting stronger vertical divergence, is observed due to a mismatch in the \textit{x–y} correlation.
	 }
      \label{fig:98_wiresim}
 \end{figure}

\begin{figure}[htb]
 \centering
	 \includegraphics[width=\linewidth]{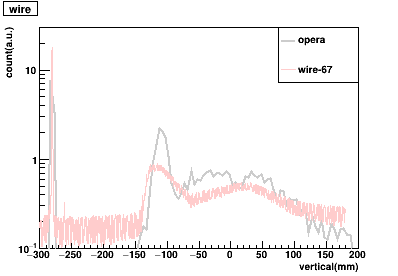}
	 \caption{Comparison with simulation and real data $\#$67 as in Fig.~\ref{fig:DC1_wire195}.
     A notable feature is that the measured beam-profile distribution for setting 98, where the \textit{x-y} coupling has the opposite sign, exhibits a shape quite similar to that obtained under the free-space condition (67).
	 }
      \label{fig:67_wiresim}
 \end{figure}

Figures~\ref{fig:95_wiresim},~\ref{fig:98_wiresim} and ~\ref{fig:67_wiresim} compare the experimental data with the corresponding simulation results obtained using the rotated-quadrupole settings listed in Table~\ref{tab:setQ123}.
Each simulation was performed with the quadrupole strengths fixed to the values shown, enabling a direct, condition-by-condition comparison with the measured beam distributions.

$\#$95 data as in fig.~\ref{fig:95_wiresim} ensures sufficient overlap with the solenoidal acceptance,
allowing more than three turns to be achieved in the actual apparatus.
Comparison between the simulation and the measured data for case $\#$98 is shown in Fig.~\ref{fig:DC1_wire195}.
Due to a mismatch in the \textit{x–y} correlation, a broader vertical distribution is observed, reflecting a stronger divergence in the vertical direction.

These measurements confirm that the strength and sign of the \textit{x-y}-coupled correlations ,~generated by the rotating-quadrupole system, directly determine the degree to which the spiral trajectory remains confined inside the solenoidal acceptance.

A notable feature is that the measured beam-profile distribution for setting 98, where the \textit{x-y} coupling has the opposite sign, exhibits a shape quite similar to that obtained under the free-space condition (67).
The simulations show the same trend: the resulting phase-space structures for 67 and 98 are more similar to each other than to the nominally well-coupled condition (95).
This behavior indicates that the sign and strength of the \textit{x-y} coupling strongly influence the effective phase-space orientation at injection.
This completes the verification of the spiral-injection principle using a DC electron beam.
The simulated slopes agree with the measurements within 5–6°, and the resulting injection trajectories are consistent with the observed number of turns.
This confirms that the transport-line model accurately captures how variations in the multivariate correlation parameters map onto the injection phase space.
Thus, the sensitivity of the injection to the \textit{x-y} correlation structure is quantitatively validated both experimentally and through model–measurement agreement.

For settings close to ID 95, the beam remains confined within a narrow vertical region over three turns.
Even though the ideal solution of fig.~\ref{fig:PhaseInj} was not fully realized,
the experimentally reconstructed phase space remains aligned with
the dominant feature direction in the $r$–$\mathrm{vertical}$–$\theta$ representation, which we will discuss it in the next section.



\subsection{Reconstruction and Interpretation of the Injection Phase Space}\label{sec:discussion1}

\begin{figure}[htb]
 \centering
	 \includegraphics[width=\linewidth]{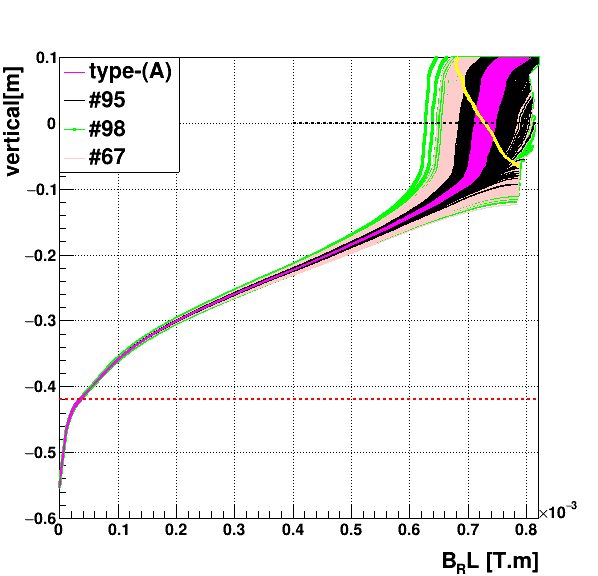}
	 \caption{Tracking results of 1000 particles for the ideal case (type-A) together with other beamline settings $\#$95, $\#$98, and $\#$67.~The horizontal axis represents <$B_{R}L$>, and the vertical axis shows the vertical position along the solenoidal axis.
The lower edge of the external iron yoke is indicated by the red dashed line at $\mathrm{vertical}=-0.42$~m, and the solenoid center is defined as $\mathrm{vertical}=0$.
	 }
      \label{fig:Ver_BrL_95_98_67_typ-A}
 \end{figure}

Figure~\ref{fig:Ver_BrL_95_98_67_typ-A} show tracking results of 1000 particles for the ideal case (type-A in Table~\ref{tab:twiss}, Fig.~\ref{fig:risouPhase}) together with other beamline settings $\#$95, $\#$98, and $\#$67, based on the actual beamline parameters in Table~\ref{tab:setQ123}.~The vertical positions along the solenoidal axis are shown as function of $\langle B_{R}L \rangle$.
The lower edge of the external iron yoke is indicated by the red dashed line at $\mathrm{vertical}=-0.42$~m, and the solenoid center is defined as $\mathrm{vertical}=0$.

The simulation results indicate that the differences in the \textit{x-y} coupling become pronounced in the region above $\mathrm{vertical} >$ -0.2~m.
For the settings with incorrect \textit{x-y} coupling ($\#$67 and $\#$98), the beam distribution rapidly expands, whereas for $\#$95 (black line), the expansion is 
suppressed.

\begin{figure}[htb]
 \centering
	 \includegraphics[width=0.8\linewidth]{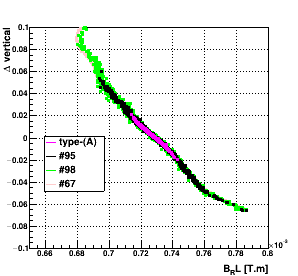}
	 \caption{An enlarged view of this yellow region is shown in Fig.~\ref{fig:slicez_BrL}.
From these figures, it is evident that $\langle B_{R}L \rangle$ and the vertical coordinate are almost linearly uncorrelated, and that the spread in $\langle B_{R}L \rangle$—and hence the beam size—directly reflects the quality of the \textit{x-y} coupling.
	 }
      \label{fig:slicez_BrL}
 \end{figure}

The yellow curve represents a slice of the distribution at the time when the central trajectory reaches $\mathrm{vertical}=0$.
An enlarged view of this yellow region is shown in Fig.~\ref{fig:slicez_BrL}.
From these figures, it is evident that $\langle B_{R}L \rangle$ and the vertical coordinate are negatively correlated, and that the spread in $\langle B_{R}L \rangle$—and hence the beam size—directly reflects the quality of the \textit{x–y} coupling.



Although the beamline monitors provide only transverse beam profiles, the preceding section has established that the transport line is sufficiently well understood to allow reconstruction of the full four-dimensional phase space.
By comparing the reconstructed phase space at the injection point for several distinct beamline settings, we relate these differences to the variation in the number of spiral turns achieved inside the storage volume.
A detailed discussion of this connection is presented in the next section.

\subsubsection{Simulation-Based Phase Space Evaluation}

\begin{figure}[htb]
 \centering
	 \includegraphics[width=\linewidth]{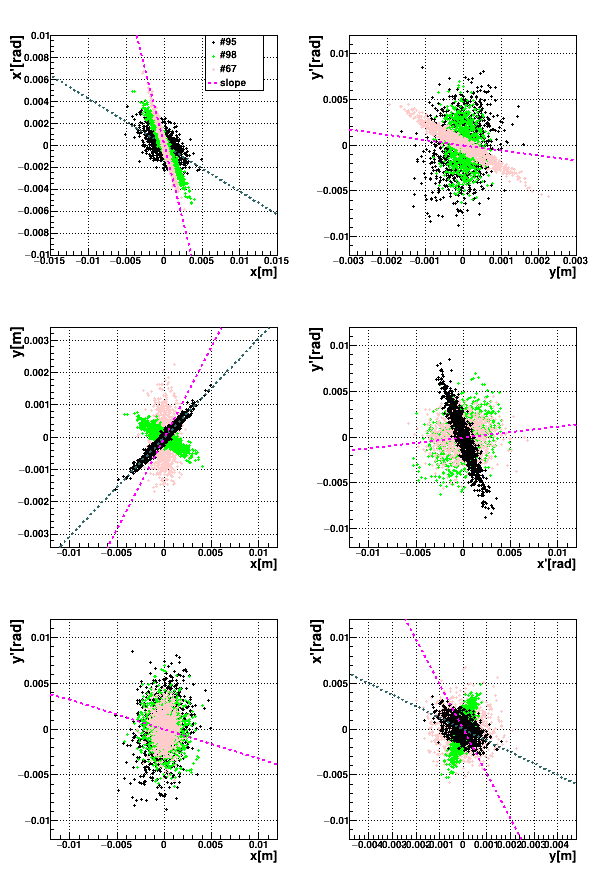}
	 \caption{ Four–dimensional phase–space ($x$, $x'$, $y$, $y'$) calculations at the
injection point for the case of $\#$95,98,67. Pink dashed line depicts slopes as explained in fig.~\ref{fig:ribbon_soukan} 
	}
 \label{fig:SixPhaseData}
 \end{figure}

In this section, the phase space is evaluated using simulations constrained by the experimentally determined beamline parameters.
Figure~\ref{fig:SixPhaseData} shows the reconstructed four-dimensional phase-space distributions corresponding to cases $\#$95, $\#$98, and $\#$67. For comparison, the slopes obtained from the ribbon-beam analysis in Fig.~\ref{fig:ribbon_soukan} are indicated by pink dashed lines.

In case $\#$95, the slopes in the \textit{x-y} and \textit{y-x'} projections differ from the target correlation by about 10°, while the \textit{x-x'} projection exhibits the expected converging slope.
In contrast, cases $\#$98 and $\#$67 do not exhibit the correct correlation directions.
In particular, the \textit{x-y} projection shows that $\#$98 has a slope opposite to that of $\#$95, whereas $\#$67 shows essentially no \textit{x-y} coupling at all.

Figure~\ref{fig:SixPhaseData} indicates that suppressing the $\mathrm{vertical}$ spread of the three-dimensional spiral trajectories inside the storage volume requires a specific prioritization in how the four-dimensional phase space is adjusted. 
Crucially, the direction of broadening matches the direction predicted by the model when $x$-$y'$ and $y$-$x'$deviate from the ideal values,~yet the ensemble still reaches the storage region for approximately 3–4 turns.

\begin{figure}[htb]
 \centering
	 \includegraphics[width=\linewidth]{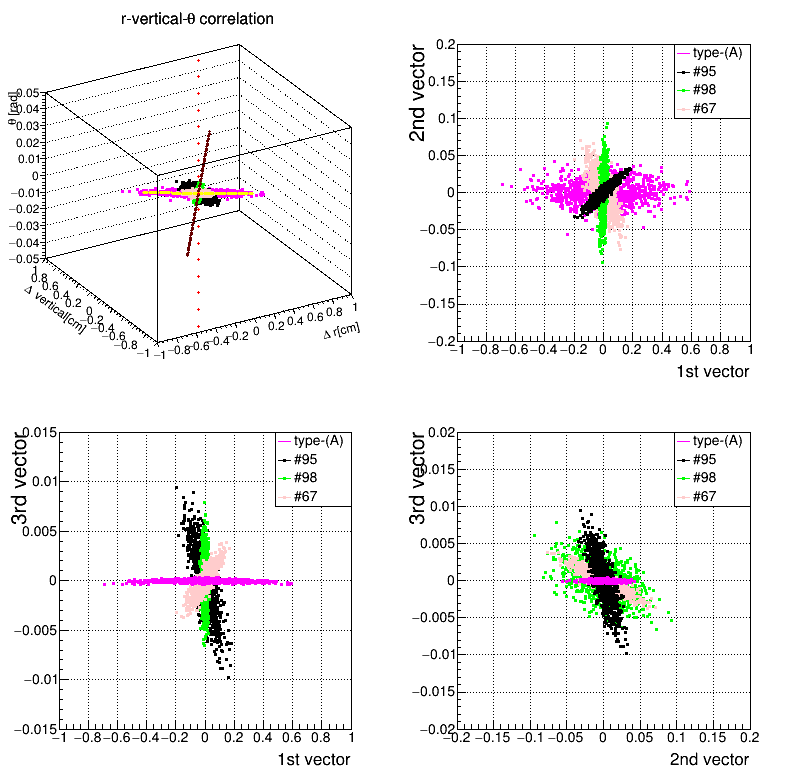}
	 \caption{
     The calculated $\Delta r$-$\Delta vertical$-$\Delta \theta$~phase-space distributions derived from the beamline settings for cases $\#$67, $\#$95, and $\#$98 are shown together with the $\Delta r$-$\Delta vertical$-$\Delta theta$~distribution corresponding to Type-(A) in Fig.~\ref{fig:r-z-thetaSim_risou_5107_516}. Singular value decomposition (SVD) performed on these distributions is summarized in Table~\ref{tab:eigenTab3}.As described in Eq.~\ref{eq:eigenVec2}, the vectors $\vec{X}$~are obtained for each case, and the angle $\delta$ relative to Type-(A) is also evaluated. Case $\#$95 is oriented close to Type-(A), whereas the others exhibit significantly different orientations.
	 }
      \label{fig:rztheta_wire95_soukan}
 \end{figure}

To substantiate this inference, Fig.~\ref{fig:rztheta_wire95_soukan} presents the same data expressed in the laboratory-frame $r$-$\mathrm{vertical}$-$\theta$~representation introduced earlier. 
The corresponding eigenvalues and eigenvectors obtained from this representation are summarized in Table~\ref{tab:eigenTab3}.

\begin{table}[hbt]
\centering
\caption{Eigen values of $\#$95,98 and 67 }
 \begin{tabular}{lllll}
\toprule
	item &  type-(A) & $\#$95 & $\#$98 & $\#$67  \\
    $\epsilon_1$   & 2.73  & $7.76\times10^{-1}$ & $3.95\times10^{-1}$ & $7.23\times10^{-1}$\\
    $\epsilon_2$   & $2.04\times10^{-1}$ & $5.49\times10^{-2}$ & $1.47\times10^{-1}$ & $2.17\times10^{-1}$\\
    $\epsilon_3$   & $1.63\times10^{-3}$ & $2.61\times 10^{-2}$ & $2.49\times10^{-2}$ &$5.15\times10^{-3}$\\
    $\delta$ & 1 &  0.984&  0.132& 0.801  \\
     \bottomrule
\end{tabular}
\label{tab:eigenTab3}
\end{table}

We define the angular deviation of the principal component vector from the ideal orientation $\delta$ as
\begin{equation}
   \delta=\frac{\vec{X}_0\cdot\vec{X}_i}{|\vec{X}_0||\vec{X}_i|}, \\ 
    \label{eq:eigenVec2} 
\end{equation}
here, $\vec{X}_0$ is a vector of type-(A) and $\vec{X}_i$ is a vector of $\#$95 or others expressed as
\begin{equation}
\vec{X}=\epsilon_1\vec{e}_1+\epsilon_2\vec{e}_2+\epsilon_3\vec{e}_3. \\
\label{eq:eigenVec}
\end{equation}

As we already discussed in Fig.~\ref{fig:r-z-thetaSim}, the model predicts that the injection performance is controlled by a dominant direction in the ($r$, $\mathrm{vertical}$, $\theta$) space, corresponding to the largest singular vector of the tSVD analysis.
When the injected phase space is aligned with this feature direction, the injected beam overlaps well with the solenoidal acceptance, and the number of turns increases.

\subsubsection{Implication for the beam injection design}\label{sec:x-ydiscussion}
When the \textit{x–y} correlation is properly matched, the orientation of the principal component vector in the ($r$, $\mathrm{vertical}$, $\theta$) phase space also becomes aligned.
The good performance observed in case $\#$95 can be attributed to the fact that the orientation of the principal axis is close to the ideal one, the eigenvalue of the first principal component is small, and the coupling to the second component is weak.
In other words, because the transverse beam cross section exhibits an appropriate \textit{x–y} correlation and a small beam size, the impact of a non-ideal momentum spread is effectively suppressed.
This result indicates that beam shaping optimized for the spatial distribution of the magnetic field is the most critical factor.
In addition, in case $\#$95, the correlations in $x$-$x'$~and~$y$–$x'$ also show a converging tendency (although not ideal), which can be interpreted as leading to a smaller ratio between the eigenvalues of the second and first principal components.
 
This agreement demonstrates that the $r$–$\mathrm{vertical}$–$\theta$ representation provides a physically meaningful framework for interpreting the injection results, even when the beamline optics cannot fully realize the ideal \textit{x-y} correlation.


To evaluate how deviations from the ideal correlation affect the injection performance, multiple transport-line settings were studied (IDs 67, 95, and 98).
Use of parameters shown in Tables~\ref{tab:eigenTab3}, we introduce a single parameter,
\begin{equation}
\chi=\frac{\epsilon_1}{\delta} \frac{\epsilon_2}{\epsilon_1} =\frac{\epsilon_2}{\delta}  \\
\label{eq:eigenVec3}
\end{equation}

 which represents the combined effect of misalignment and residual coupling.
We examined the correlation between this parameter and the beam size (or beam spread) at a given observation point.

\begin{figure}[htb]
 \centering
	
	 \includegraphics[width=0.8\linewidth]{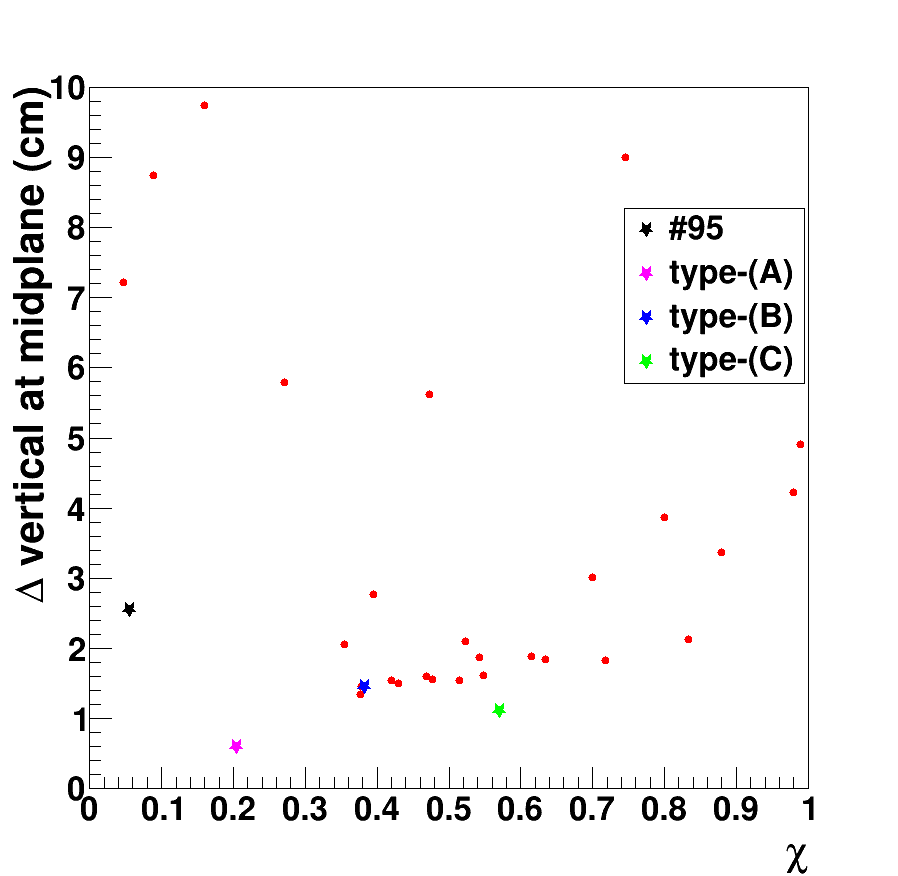}
	 \caption{
     The parameter $\chi$~defined in Eq.~\ref{eq:eigenVec3}, is shown together with the vertical beam spread obtained from a time slice in the storage chamber. The star symbols correspond to the cases indicated in the legend, while the red points represent additional cases evaluated in the same manner. A general tendency is observed that smaller $\chi$~values correspond to reduced vertical beam spread. Although the principal component vector for case $\#$95 is not perfectly aligned with Type-(A), the second eigenvalue is sufficiently small, resulting in a small $\chi$~value.
	 }
      \label{fig:kime}
 \end{figure}

Although the correlation is not strictly monotonic,
the parameter $\chi=\delta\epsilon_2$ provides a meaningful measure
to characterize the lower bound of the beam spread achievable at a given point.

In particular, both the ideal solution obtained from calculations and
the experimentally optimized case ($\#$195) are located in the region of
small $\chi$, supporting the validity of this parameter
as a practical figure of merit for injection tuning.



\section{Discussion}
Traditionally, the storage of charged particle beams has been limited to two options:
either confining high-energy particles in large storage rings with circumferences of several kilometers, or decelerating particles to near rest and capturing them in traps.
The results of this study demonstrate a third option, in which charged particle beams with low to intermediate momentum can be stored in a compact ring with a diameter of less than 1 m while maintaining relativistic energy.

When storing a beam inside a single solenoid magnet, the matching of the principal component of the beam phase space and the control of X–Y coupling at the injection point must be defined in terms of the spatial distribution of the fringe magnetic field near the injection region.
In this respect, the beam control method demonstrated in this study is based on a control concept that fundamentally differs from conventional beam transport schemes, which assume phase-space design around a reference trajectory propagating through free space.

When a beam with finite emittance—characterized by finite transverse size and momentum spread—passes through a magnetic field with a non-uniform spatial distribution, the beam phase-space distribution must be adapted to the magnetic field profile.
Rather than deriving the phase space from local magnetic field distributions, this study shows that an ideal injection condition is achieved when the integrated magnetic field BrL experienced by individual charged particles around the injection trajectory matches the integrated magnetic field BrL. experienced by the reference trajectory.
The corresponding condition is obtained using a reverse-tracking method, which yields the optimal four-dimensional correlated phase-space distribution for a given external magnetic field configuration.

The first technical core of this experimental demonstration lies in the evaluation and control of beam phase space under conditions where the reference trajectory is not a simple straight line.
In general, phase-space correlations defined in the beam coordinate system fully characterize beam shaping. However, in the present injection scheme, the reference trajectory is curved, and therefore the use of the $r$–$vertical$–$\theta$ phase-space representation in the laboratory coordinate system enables precise evaluation and control of the beam phase space.
This feature arises from the presence of an external magnetic field that defines a non-linear reference trajectory.
\section{Summary}
This paper reports a demonstration experiment of three-dimensional spiral injection into a compact ring with a circumference of less than 20 cm, with particular emphasis on the design, control, and operational results of strongly X–Y coupled beams.
An 80~keV DC electron beam was injected from an electron gun into a solenoidal magnetic field along a three-dimensional spiral trajectory. Multi-turn beam orbits near the magnet center were evaluated qualitatively by optical visualization and quantitatively using wire scanners, and the results were compared with orbit simulations.

To reach the final results, we present a structured discussion of the design concept of the demonstration beamline, the outline of the experimental setup, representative examples of tuning X–Y coupled beams, and the achievable range of X–Y coupling in this beamline. We further clarify the prioritization among multiple parameters required to realize the target X–Y coupling and validate this strategy experimentally.
In this sense, the present paper serves as a technical record of how the design concept and the implementation were reconciled with practical constraints, and provides a useful guideline for approaching design targets even when realistic limitations—such as restricted footprint and installation space—prevent full satisfaction of the ideal requirements.

While this paper focuses on three-dimensional spiral-orbit tuning using a DC beam, the ultimate goal of the beamline is to establish a technique for slicing the DC beam into pulses and storing it in the storage region for a finite duration. Results on pulsed-beam injection and storage will be reported elsewhere.

\section{Acknowledgment}
We thank H.~Hirayama, and K.~Oda for their essential contributions during the initial setup and early commissioning of the experiment, although they were not involved in the data-taking period relevant to this manuscript.
The authors gratefully acknowledge, H. Hisamatsu, H. Someya, T. Suwada, Y. Okayasu, and Y. Yano for providing the equipment used in this study. We further acknowledge T. Ushiku of Next Create Service Co.,Ltd. for magnet fabrication. We also acknowledge the members of Futaba Kogyo Co., Ltd. for their long-term technical support in the installation and operation of the experimental equipment. We thank the KEK Mechanical Engineering Center for their technical assistance and the KEK Accelerator Injector–LINAC Group for their general support.

We express our sincere gratitude to K. Oide, whose discussions provided the initial conceptual inspiration for this work.
We also thank colleagues at KEK Institute of Particle and Nuclear Studies, including N. Saito and T. Mibe, for their long-standing support.

This work was supported by JSPS KAKENHI Grant Nos. 26287055, 19H00673, 22K14061, and 23KJ0590.

\newpage
\clearpage
\appendix 
\section{Appendix Section}\label{app1}
\subsection{Phase-Space Representation and $\Sigma$-Matrix Formalism}
\label{sec:R-mat}
We first present the fundamental expressions for describing the phase space, introducing the $\Sigma$-matrix and its ten independent components.
\begin{eqnarray}
	\Sigma&=& 
            \begin{bmatrix}
		    \langle xx\rangle   &\langle xx'\rangle  &\langle xy\rangle   &\langle xy'\rangle   \\
		    \langle x'x\rangle  &\langle x'x'\rangle &\langle x'y\rangle &\langle x'y'\rangle   \\
		    \langle yx\rangle   &\langle yx'\rangle  &\langle yy\rangle   &\langle yy'\rangle   \\
		    \langle y'x\rangle  &\langle y'x'\rangle &\langle y'y\rangle  &\langle y'y'\rangle   \\
	     \end{bmatrix}  
    \label{eq:sigma-1}
\end{eqnarray}
Here, we take the beam coordinate system; horizontal distribution around the reference orbit as the $x$ axis and the vertical distribution as the $y$ axis, the four-dimensional transverse phase space 
($x$,$x'$,$y$,$y'$). And the $\Sigma$-matrix contains ten independent variables:$\langle xx\rangle$, $\langle xx'\rangle$, $\langle x'x'\rangle$, $\langle yy\rangle$,   $\langle yy'\rangle$, $\langle y'y'\rangle$, $\langle xy\rangle$, $\langle xy'\rangle$, $\langle x'y\rangle$ and $\langle x'y'\rangle$. Latter four variables are coupled in $x$ and $y$ axis components.

In the case of latter four components are all zero, the correlations between the $x$ and $y$ directions are negligible, 
$\Sigma$-matrix~can be written as a block-diagonal matrix consisting of the $2\times2$~submatrices $\sigma_x$~and~$\sigma_y$, i.e.,

\begin{eqnarray}
	\Sigma_0&=& 
            \begin{bmatrix}
		   \sigma_x   &  0  \\
		    0        &  \sigma_y
	     \end{bmatrix} \\ \nonumber
	\sigma_{x(y)}&=& {\epsilon_{x(y)}}^2
            \begin{bmatrix}
		    \beta_{x(y)}   & -\alpha_{x(y)}  \\
		    -\alpha_{x(y)} & \gamma_{x(y)}
	 \end{bmatrix}
    \label{eq:sigma-2}
\end{eqnarray}

Here, $\epsilon_{x(y)}$~denotes the emittance, which is a conserved quantity.
The parameters $\alpha_{x(y)}, \beta_{x(y)}$ and $\gamma_{x(y)}$~are referred to as the Twiss parameters; their detailed definitions can be found in standard textbooks.

In case of \textit{x-y} coupling, $\Sigma$-matrix , can be expressed in terms of the uncoupled sigma matrix $\Sigma_0$ 
 and the so-called $R$-matrix, which introduces the coupling, as follows:
\vspace{-2mm}
\begin{equation}
	\Sigma_{xy}=R^{-1}\Sigma_0(R^{-1})^{t}
\label{eq:sigma-3}
\end{equation}

$R$-matrix is expressed as
\vspace{-2mm}
\begin{equation}
	R=
            \begin{bmatrix}
		    \mu   & 0    & -r_4 & r_2  \\
			0 & \mu  &  r_3 & r_1  \\ 
		     r_1 & r_2  &\mu   & 0    \\
		      r_3 & r_4  &    0 & \mu  \\ 
	 \end{bmatrix}
\label{eq:r_mat}
\end{equation}

The ten independent variables in Eq.~\ref{eq:sigma-1} can be parameterized by the ten Twiss-parameters shown in Eqs.~\ref{eq:sigma-2} and \ref{eq:r_mat}, namely~$\epsilon_{x,(y)}$, $\alpha_{x(y)}$, $\beta_{x(y)}$, $r_1$,$r_2$,$r_3$,$r_4$.
In other words, if one could freely choose these ten parameters as determined by the transport line, an arbitrary~$\Sigma$~matrix could, in principle, be constructed.

\section{Beamline Description and Transfer-Matrix Formalism}\label{sec:SetupDetails}
In this appendix, we present the transfer matrices associated with the rotated quadrupoles in the transport line used in the demonstration experiment, and derive the transfer matrices from the beamline to the two beam diagnostic points installed along the line.
Details are find in these papers:~\cite{rehman:ipac2021-mopab162}~\cite{RehmanLINAC18},~\cite{RehmanIQBS2019},~\cite{RehmanThesis},~\cite{Hira-MT27},~\cite{matsu_master}.

\begin{figure}[htb]
 \centering
	 \includegraphics[width=\linewidth]{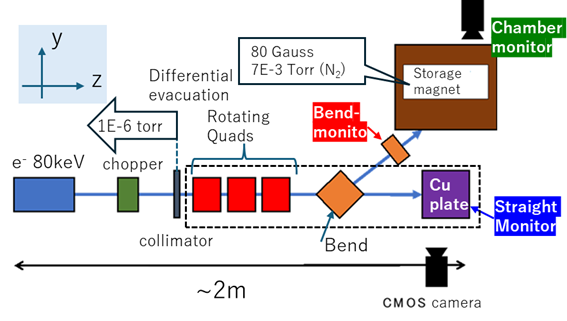}
	 \caption{Outline of SITE. 
	  Monitor-1 and Monitor-3 measure X-Y coupling and 3-D spiral trajectories.
	 }
      \label{fig:BeamLineFig}
 \end{figure}

\begin{figure}[htb]
 \centering
	 \includegraphics[width=\linewidth]{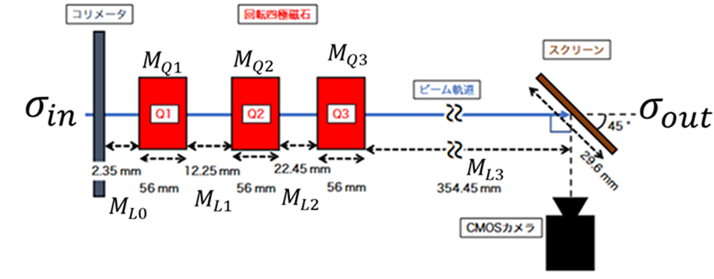}
	 \caption{Zoomup view rotating quadrupoles positions and straight diagnastic line.
	 }
      \label{fig:BeamLineFigZoom}
 \end{figure}

In this section, we adopt a beam coordinate system in which the beam axis is taken as the normal ($z$ axis), 
with the vertical ($y$) and horizontal ($x$) directions defined on the transverse plane.
Figure~\ref{fig:BeamLineFig} and Fig.~\ref{fig:BeamLineFigZoom} show three quadrupole magnets installed between the beamline collimator and the bending magnet.

Each quadrupole is equipped with a mechanism that allows arbitrary rotation about the beam axis, 
thereby introducing coupling in the beam motion in horizontal ($x$ axis) and vertical ($y$ axis) in~\ref{fig:BeamLineFig}.

The transverse beam phase space projected onto the $x–y$ plane is tuned using the three rotating quadrupoles, 
which are located between the bending magnets as shown in Fig.~\ref{fig:BeamLineFigZoom}.


 \begin{figure}[htb]
 \centering
	 \includegraphics[width=\linewidth]{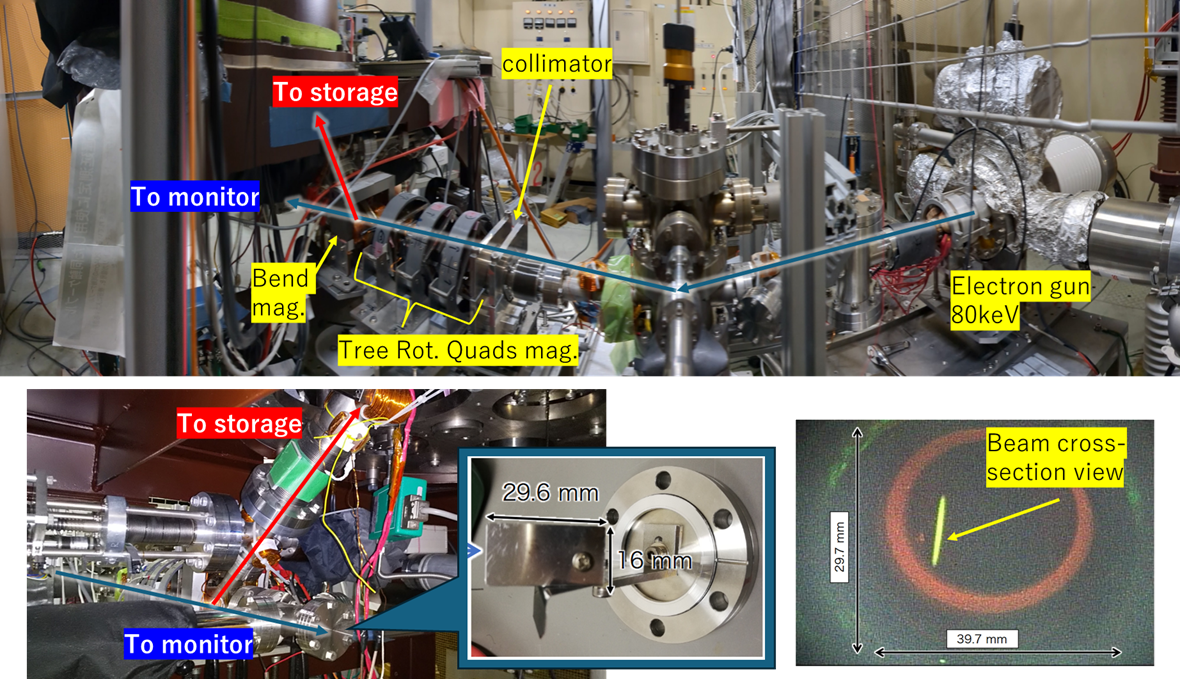}
	 \caption{Beam line and the cross-section monitor at the end of the straight section.
	 }
      \label{fig:StraightMon}
 \end{figure}

 The transfer matrix of a \textbf{single} quadrupole magnet rotated by an angle~$\phi_{Q}$~about the beam axis can be explicitly expressed as follows:
\vspace{-2mm}
\begin{equation}
	R_{\phi_Q}=
            \begin{bmatrix}
		    \cos\phi_Q   & 0    & \sin\phi_Q & 0  \\
			0 & \cos\phi  &  0 & \sin\phi_Q  \\ 
		     -\sin\phi_Q & 0  & \cos\phi_Q   & 0    \\
		      0 & -\sin\phi_Q  &    0 & \cos\phi_Q  \\ 
	 \end{bmatrix}
\label{eq:R_rot}
\end{equation}

Substituting $r_1=r_4=\cos(\phi_Q)$,~and~$r_2=r_3=0$~into Eq.~\ref{eq:r_mat} yields this form.
If the transfer matrix of a normal (non-rotated) quadrupole is denoted by 
$N_Q$, the transfer matrix of a quadrupole magnet rotated by an angle 
$\phi_{Q}$can be written as follows:

\begin{equation}
	M_{Q}={R_{\phi_Q}}^{-1}~N_{F(D)}~R_{\phi_Q} \\ 
\label{eq:mat_rotQ}
\end{equation}
Here, a normal quadrupole magnet (focusing in the $x$~direction and defocusing in the $y$~direction) is represented by the following expression:
\vspace{-2mm}
\begin{equation}
	N_{F}=
            \begin{bmatrix}
		    a1 &a2 & 0  & 0     \\
		     a3 & a1  & 0 & 0   \\ 
            0  & 0 & b1 & b2     \\
		      0 & 0  &b3 & b1 \\ 
	 \end{bmatrix}
\label{eq:NormalQ1}
\end{equation}
for $k_Q >$0, and here,
\begin{eqnarray}
     a1&=&\cos(\sqrt{k_Q}L_Q)\\ \nonumber
     a2&=&\frac{\sin(\sqrt{k_Q)}L_Q)}{\sqrt{k_Q}} \\ \nonumber
   a3&=&-\sqrt{k_Q}\sin(\sqrt{k_Q}L_Q) \\ \nonumber
   b1&=&\cosh(\sqrt{k_Q}L_Q) \\ \nonumber
   b2&=&\frac{\sinh(\sqrt{k_Q}L_Q)}{\sqrt{k_Q}}\\ \nonumber
   b3&=&\sqrt{k_Q}\sinh(\sqrt{k_Q}L_Q).
\label{eq:Q_comp}
\end{eqnarray}

  Similarly, in case of $k_Q<$0, set $k_Q=k_Q\times-1$ to substitute numerical values into Eq.~\ref{eq:Q_comp} and then,
\begin{equation}
	N_{D}=
            \begin{bmatrix}
		    b1 &b2 & 0  & 0     \\
		     b3 & b1  & 0 & 0   \\ 
            0  & 0 & a1 & a2     \\
		      0 & 0  &a3 & a1 \\ 
	 \end{bmatrix}
\label{eq:NormalQ1}
\end{equation}

Table~\ref{tab:SITE2} lists the parameters $k_Q$,~$L_Q$ and $\phi_Q$ of the three rotating quadrupoles.

\begin{table}[!hbt]
   \centering
   \caption{ Q1,~Q2~and~ Q3 parametersin Fig.~\ref{fig:BeamLineFigZoom} and \ref{fig:BeamLineFigZoomBend}}
   \begin{tabular}{lcc}
       \toprule
	   symbol& values &note \\
       \midrule
	  $L_Q$   & 56~mm      & effective length     \\
	   $k_Q$   & -120~$\sim$~120[T/m]  & k-values  \\
	   $\phi_Q$   & -45~$\sim$~450[degree]  & Rotating angle  \\
       \bottomrule
   \end{tabular}
   \label{tab:SITE2}
\end{table}

Table~\ref{tab:SITE1} lists the parameters of the three rotating quadrupol the drift spaces from the collimator to the downstream end of the straight section.
In the present discussion, the bending magnets are treated as drift spaces.

\begin{table}[!hbt]
   \centering
	\caption{Free space length in Fig.~\ref{fig:BeamLineFigZoom} and \ref{fig:BeamLineFigZoomBend}}
   \begin{tabular}{lcc}
       \toprule
	   symbol& values &note \\
       \midrule
	  $L_0$   & 2.35~mm    &  collimeter and $Q_1$  \\ 
	  $L_1$   & 12.25~mm   &  $Q_1$ and $Q_2$ \\ 
	  $L_2$   & 22.45~mm   &  $Q_2$ and $Q_3$     \\ 
	  $L_3$   & 354.45~mm  &  $Q_3$ and screen     \\ 
	  $L_4$   & 10.405~mm  &  $Q_3$ and bend     \\ 
	  $L_5$   & 111.682~mm  &  bend and screen     \\ 
       \bottomrule
   \end{tabular}
   \label{tab:SITE1}
\end{table}

Transfer matrix of from the collimator to the cupper screen is expressed as:
\vspace{-2mm}
\begin{equation}
	M=M_{L3}\cdot~M_{Q3}\cdot~M_{L2}\cdot~M_{Q2}\cdot~M_{L1}\cdot~M_{Q1}\cdot~M_{L0}
\label{eq:mat_straight}
\end{equation}
Here, 
$L_0$,$L_1$,, denote the transfer matrices of the drift spaces, and 
$M_{Q1}$~represents the transfer matrix of the $Q_1$ quadrupole magnet alone.
Figure~\ref{fig:BeamLineFigZoomBend} depicts another called "Bend-monitor" set at the injection point.
 \begin{figure}[htb]
 \centering
	 \includegraphics[width=\linewidth]{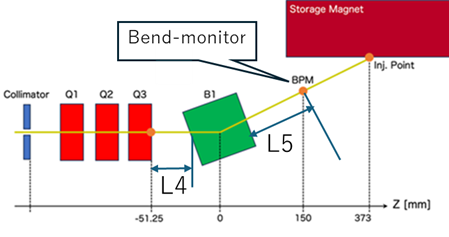}
	 \caption{Zoomup view rotating quadrupoles positions and straight diagnastic line.
	 }
      \label{fig:BeamLineFigZoomBend}
 \end{figure}

 \begin{figure}[htb]
 \centering
	 \includegraphics[width=\linewidth]{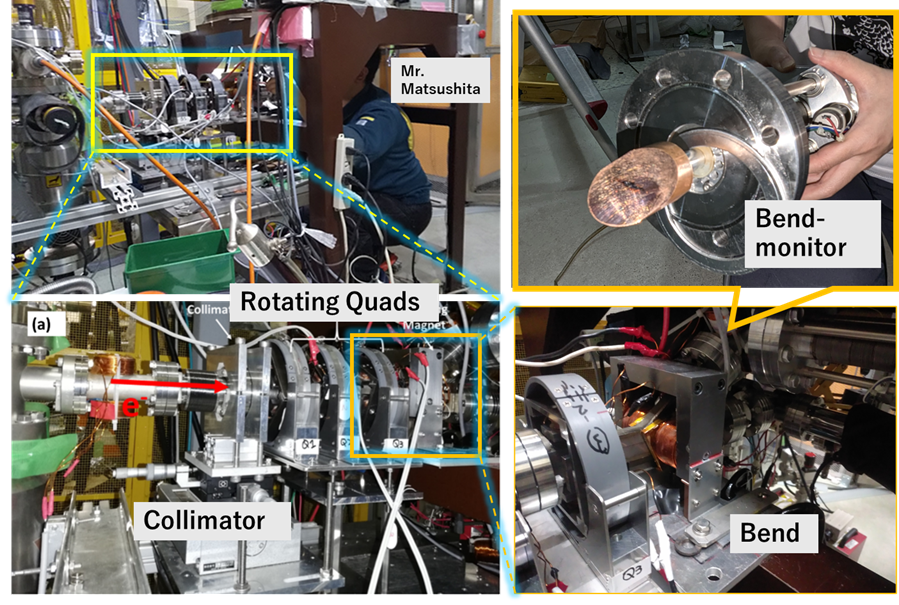}
	 \caption{Left above:Down stream of the beam line. Left bottom: Collimator, three rotating quadrupoles and bend magnet from the left. Right bottom:Zoom up view of the bend magnet to transport beam 45 degrees upward towards bean injection. Right above: zoom up view of bend monitor, which is well polished copper. The black burnt area in the center is a trace left over from continuous beam measurements. 
	 }
      \label{fig:BeamLineFigZoomBend}
 \end{figure}

 \begin{figure}[htb]
 \centering
	 \includegraphics[width=0.9\linewidth]{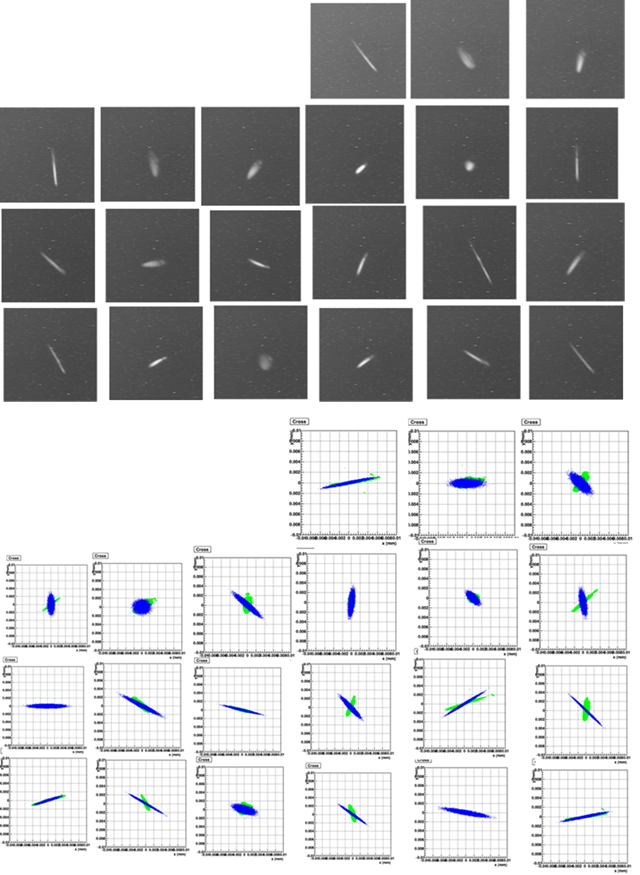}
	 \caption{
     Up:Beam cross-section views at Monitor-2. Bot-
tom: Comparison of beam cross-sections by changing beam
line settings. These are correspond to fig~\ref{fig:fig13}~data set.
	 }
      \label{fig:BendView}
 \end{figure}
Figure~\ref{fig:BendView} shows the beam cross-section at the injection point downstream of the bending magnet, together with a comparison to the simulation.

\section{Reconstruction of the Full $\Sigma_0$-Matrix from Transverse \textit{x-y} Measurements}\label{sec:Sigma-measure}

Pictures of test beam line to demonstrate a new concept of beam injection scheme are shown in the main text of Fig.~\ref{fig:SITEAll},~\ref{fig:BeamLinePict}. 
Some major parameters are listed in Table~\ref{tab:Tab1} as well as parameters for actual experiment.

More details will be discussed in thie section.
At the end of the straight section in Fig.~\ref{fig:StraightMon}, there is an copper plate to detect cross-section view of beam. Example pictures named "Non X-Y coupled" and "X-Y coupled" beams. Three rotating quadruple magnets are utilized to try several types of X-Y coupling.

Figure~\ref{fig:fig13} in the main text depicts beam cross-section views with twenty different quadruple settings to estimate beam space parameters.
In general, \textit{x-y} coupled beam is expressed in ten parameters ($\epsilon_x$, $\epsilon_y$, $\alpha_x$,$\alpha_y$,$\beta_x$,$\beta_y$, and $r_1, r_2, r_3$ and $r_4$), and we estimated them from the measured data. Note that we treat transverse phase space only in this paper.
In principle, eight parameters except for emittance should be controlled with at least eight independent tuning knobs. 
However, there are only six tuning knob (three rotating angles, three K-values) in this beam line, because of a space limitation as in Fig.~\ref{fig:StraightMon} and ~\ref{fig:BeamLineFigZoomBend}.
Therefore, our system takes steps to prioritize and adjust the \textit{x-y} correlation to be ideal, as we show comparisons of twenty cross-section views of real and simulated beam shapes in Fig.~\ref{fig:fig13}
Note that the ideal design should have made redundant using seven rotating quadrupole magnets~\cite{Reference_3},~\cite{IPAC2023},~\cite{Iinuma:ipac2025-wepm029},~\cite{Iinuma:ipac2024-wepg46} to adjust six phase-space correlations ($x$-$x$', $y$-$y'$, $x$-$y$, $x$-$y'$ and $y$-$x'$ in the beam coordinate).

Three measured values at the straight beam monitor can be expressed as follows by use of 4-by-4 transfer matrix of $i^{th}$ setting beam parameter:$M$.

\begin{eqnarray}
	\langle xx \rangle_i 
	&=& M_{11}^{2}\langle xx \rangle_{in}+2M_{11}M_{12}\langle xx' \rangle_{in}+M_{12}^{2}\langle x'x'\rangle_{in} \\ \nonumber 
	&&+M_{13}^2 \langle y^2 \rangle + 2M_{13}M_{14}\langle yy' \rangle +M_{14}^2 \langle y'y' \rangle \\ \nonumber
	&&+2M_{11}M_{13} \langle xy \rangle + 2M_{11}M_{14}\langle xy' \rangle  \\ \nonumber
	&&+2M_{12}M_{13}\langle x'y \rangle+2M_{12}M_{14} \langle x'y' \rangle,
    \label{eq:xx}
\end{eqnarray}
here, subscripts denote each component of $M$.
\begin{eqnarray}
	\langle yy \rangle_i 
	&=& M_{31}^{2}\langle xx \rangle_{in}+2M_{31}M_{32}\langle xx' \rangle_{in}+M_{32}^{2}\langle x'x'\rangle_{in} \\ \nonumber 
	&&+M_{33}^2 \langle y^2 \rangle + 2M_{33}M_{34}\langle yy' \rangle +M_{34}^2 \langle y'y' \rangle \\ \nonumber
	&&+2M_{31}M_{33} \langle xy \rangle + 2M_{31}M_{34}\langle xy' \rangle  \\ \nonumber
	&&+2M_{32}M_{33}\langle x'y \rangle+2M_{32}M_{34} \langle x'y' \rangle
    \label{eq:yy}
\end{eqnarray}

\begin{eqnarray}
	\langle xy \rangle_i 
	&=& M_{11}M_{31}\langle xx \rangle_{in}+(M_{11}M_{32}\\ \nonumber 
	&& +M_{12}M_{31})\langle xx' \rangle_{in} \\ \nonumber 
	&& +M_{12}M_{32}\langle x'x'\rangle_{in} +M_{13}M_{33} \langle y^2 \rangle  \\  \nonumber 
	&& +(M_{13}M_{34}+M_{14}M_{33})\langle yy' \rangle +M_{14}M_{34} \langle y'y' \rangle \\ \nonumber
        &&+(M_{11}M_{33}+M_{13}M_{31}) \rangle xy \langle  \\ \nonumber 
	&&+ (M_{12}M_{33}+M_{13}M_{32})\langle xy' \rangle \\ \nonumber 
	&&+(M_{14}M_{31}+M_{11}M_{34}) \langle x'y \rangle \\ \nonumber 
	&& + (M_{12}M_{34}+M_{14}M_{32})\langle x'y'\rangle \\  \nonumber
    \label{eq:xy}
\end{eqnarray}

When we have 20 independent data set, then we have
\vspace{-2mm}
\begin{equation}
	\begin{bmatrix}
		\langle xx \rangle_{0} \\ \nonumber
		\langle yy \rangle_{0} \\ \nonumber
		\langle xy \rangle_{0} \\ \nonumber
		..... \\ \nonumber
		\langle xx \rangle_{i} \\ \nonumber
		\langle yy \rangle_{i} \\ \nonumber
		\langle xy \rangle_{i} \\ \nonumber
		..... \\ \nonumber
		\langle xx \rangle_{20} \\ \nonumber
		\langle yy \rangle_{20} \\ \nonumber
		\langle xy \rangle_{20} \\ \nonumber
	\end{bmatrix}_{E} = M_{mon}
	\begin{bmatrix}
		\langle xx \rangle\\ \nonumber
		\langle xx' \rangle\\ \nonumber
		\langle x'x' \rangle\\ \nonumber
		\langle yy \rangle\\ \nonumber
		\langle yy' \rangle \\ \nonumber
		\langle y'y' \rangle \\ \nonumber
		\langle xy \rangle\\ \nonumber
		\langle xy' \rangle \\ \nonumber
		\langle x'y \rangle \\ \nonumber
		\langle x'y' \rangle
	\end{bmatrix}_{in}  
\label{eq:mat_straight2}
\end{equation}

Obtaining $M_{mom}$ from each beam transport-line setting parameters, then $\Sigma_0$-matrix at initial point, as introduced in Equation~\ref{eq:sigma-1}, is estimated. Detailed explanations should be found in Dctrola Thesis of Dr. M.A.~Rehman~\cite{RehmanThesis}.

\begin{figure}[htb]
 \centering
	 \includegraphics[width=0.8\linewidth]{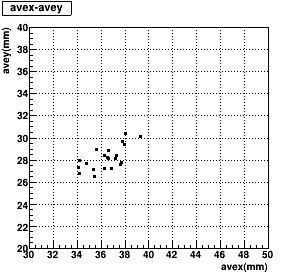}
	 \caption{Center points distributions width of fig.~\ref{fig:fig13} views. 
	 }
      \label{fig:centerZure}
 \end{figure}

Figure~\ref{fig:centerZure} shows the beam-center displacement observed for twenty different scan data sets. Misalignment of a quadrupole magnet introduces a small kick to the beam trajectory, which can be detected as a shift of the beam center at downstream beam monitors. From this figure, the quadrupole misalignment is estimated to be within 100 $\mu$m, consistent with independent measurements evaluating the reproducibility of the setup during on-site operations. This level of misalignment is sufficiently smaller than the effective magnetic-field region of the quadrupole magnet and does not cause any significant impact on beam operation.
At the end of this section, we would note that applied beam line modifications several times:Major update of the beam line since 2020.
\begin{itemize}
     \item Replacement of the straight-section monitor: 19 Oct 2021~ (Logbook No. 5, p. 25)
     \item Replacement of the bend monitor: 6 Jul 2023~(Logbook No. 5, p. 186)
     \item  Replacement of the bending magnet: before October 2020.
     \item Install of the jig for fixing the feed-through: 3 Nov 2021
    (Logbook No. 5, p. 37)~A systematic offset of approximately 10 mm was reduced to the level of the actuator precision, about 1 mm.~This improvement was largely due to a change in the measurement method, rather than the presence of the jig itself.(Logbook No.5, p.123).
\end{itemize}.
 \section{ Ribbon beam study to confirm sensitivity of \textit{x-y}~Correlations}
 \label{sec:X-Ysoukan}

\begin{figure}[htb]
 \centering
	 \includegraphics[width=\linewidth]{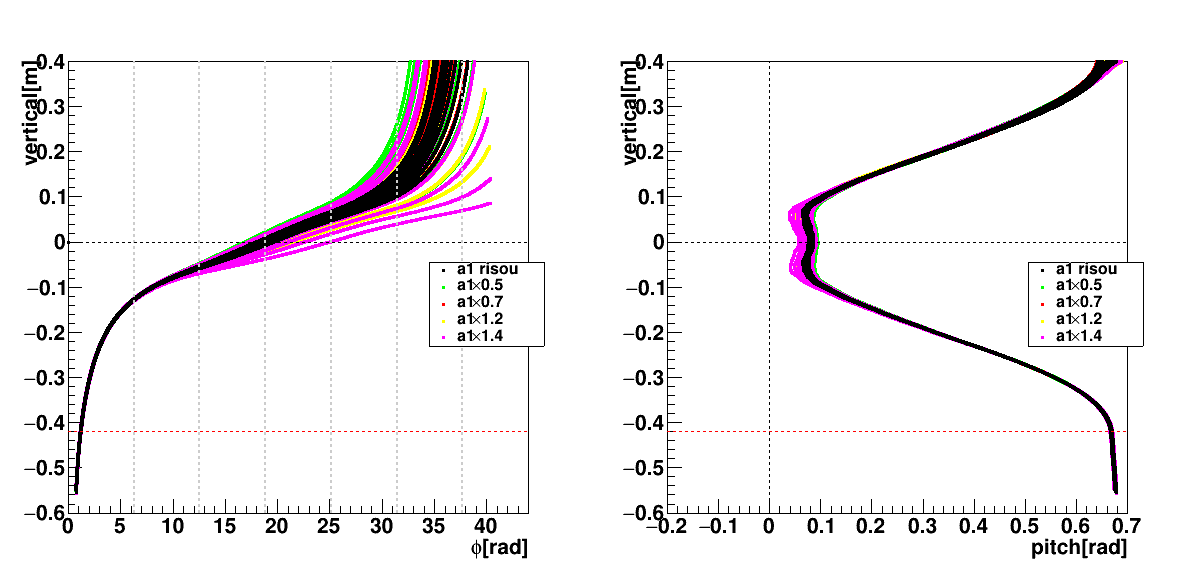}
	 \caption{Same as Fig.~\ref{fig:A1_A1Reverse} but change a1 correlations by factor 0.5~1.4. 
	 }
      \label{fig:a1_soukan}
 \end{figure}

\begin{figure}[htb]
 \centering
	 \includegraphics[width=\linewidth]{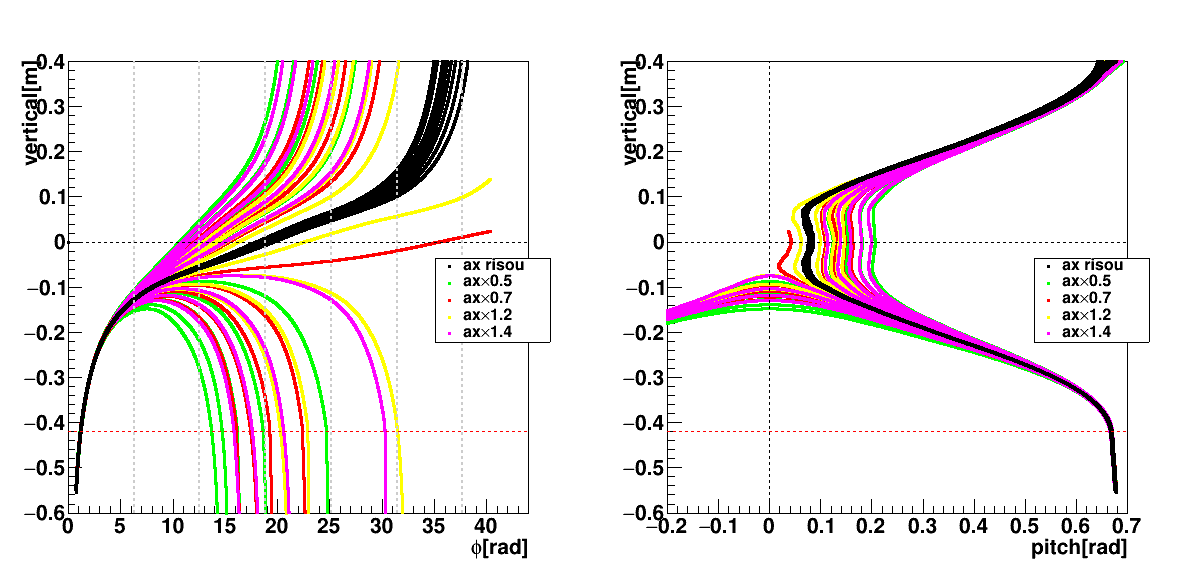}
	 \caption{Same as Fig.~\ref{fig:A1_A1Reverse} but change ax correlations by factor 0.5~1.4. 
	 }
      \label{fig:ax_soukan}
 \end{figure}

\begin{figure}[htb]
 \centering
	 \includegraphics[width=\linewidth]{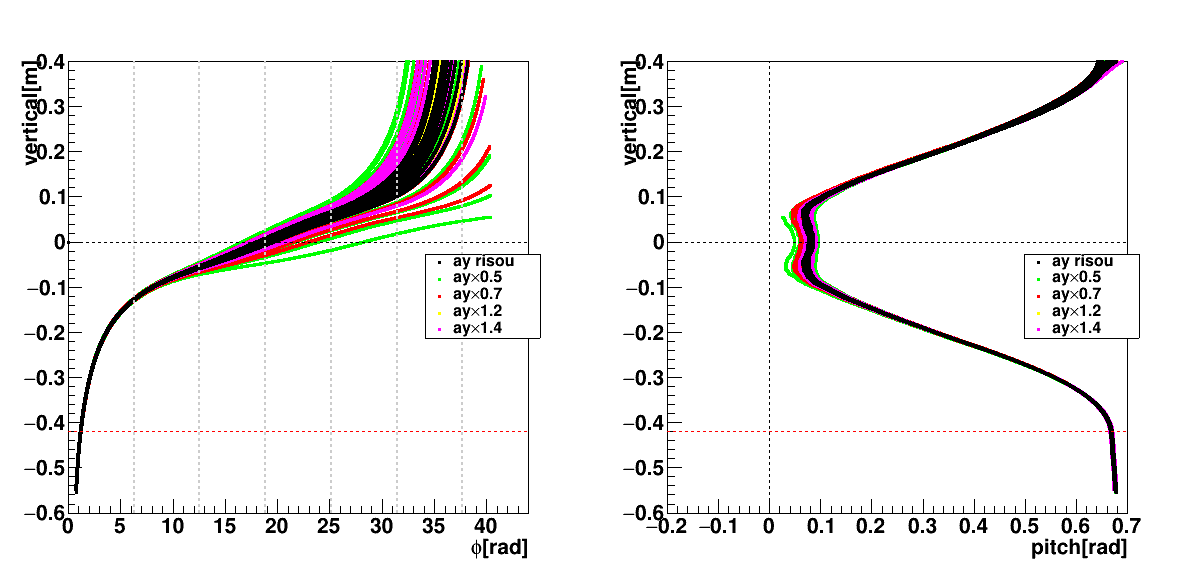}
	 \caption{Same as Fig.~\ref{fig:A1_A1Reverse} but change ay correlations by factor 0.5~1.4. 
	 }
      \label{fig:ay_soukan}
 \end{figure}

In this subsection, we clarify the relationship between the correlation parameters of the flat (ribbon) beam and the beam spread along the solenoid axis inside the storage magnet.
Among the correlation parameters listed in Table~\ref{tab:soukan}, we vary
$x-x'$, and $y-y'$ by scaling their nominal values from 0.5 to 1.4, 
and present the corresponding simulation results—distinguished by different colors—in Figs.~\ref{fig:a1_soukan}, \ref{fig:ax_soukan}, and \ref{fig:ay_soukan}, respectively.



From these comparisons, it is evident that the correlation that should be prioritized for tuning is
$x-x'$, as shown in Fig.~\ref{fig:ax_soukan}.
This is because the strong radial dependence of the radial magnetic-field component~$B_R$ significantly affects 
the rate at which the injection angle~$\theta$~decreases during injection.

Note that, for a circumference of 0.7~m per turn, an angular spread of $\Delta \theta$=0.02 rad corresponds to 
an orbit spread of approximately~0.14 m.

\end{document}